\documentclass[twocolumn,floatfix]{aastex7}
\usepackage{amsmath}
\usepackage{wasysym}     
\usepackage{graphicx}
\usepackage{amssymb}
\usepackage{dsfont}
\usepackage{commath}
\usepackage{epstopdf}
\usepackage{mathrsfs}
\usepackage{anyfontsize}
\usepackage{natbib}
\usepackage{color}
\usepackage{lipsum}
\usepackage{diagbox}

\shorttitle{Mass Transfer Stability in rpTDEs}
\shortauthors{Bandopadhyay, Coughlin, \& Nixon}
\begin{document}
\title{Repeated Tidal Interactions between Stars and Supermassive Black Holes: Mass Transfer, Stability, and Implications for Repeating Partial Tidal Disruption Events}
\author[0000-0002-5116-844X]{Ananya Bandopadhyay}
\affiliation{Department of Physics, Syracuse University, Syracuse, NY 13210, USA}
\email[show]{abandopa@syr.edu}

\author[0000-0003-3765-6401]{Eric R.~Coughlin}
\affiliation{Department of Physics, Syracuse University, Syracuse, NY 13210, USA}
\email[show]{ecoughli@syr.edu}

\author[0000-0002-2137-4146]{C.~J.~Nixon}
\affiliation{School of Physics and Astronomy, Sir William Henry Bragg Building, Woodhouse Ln., University of Leeds, Leeds LS2 9JT, UK}
\email[show]{c.j.nixon@leeds.ac.uk}

\begin{abstract}
Stars orbiting supermassive black holes can generate recurring accretion flares in repeating partial tidal disruption events (TDEs). Here we develop an efficient formalism for analyzing the time-dependent response of a star to the removal of a fraction ($\lesssim 10\%$) of its mass. This model predicts that mass loss results in a decrease in the average density of low-mass ($\lesssim 0.7 M_{\odot}$) stars. Contrarily, higher-mass stars exhibit an increase in their average density, such that the change is more pronounced for larger mass losses, and stars with masses $\sim 1.5-2 M_{\odot}$ experience the largest such increase. We predict that the final energy of the star post-mass-loss (i.e., the ``surviving core'') is effectively given by the binding energy of the original star interior to the radius from which mass is removed, i.e., the final core energy is agnostic to the process that removes the mass and -- as a corollary -- tidal heating is comparatively insignificant. We find excellent agreement between our predictions and one-dimensional Eulerian simulations of a star undergoing mass loss, and three-dimensional Lagrangian simulations of partial TDEs. We conclude that 1) partially disrupted stars are not significantly heated via tidal dissipation, 2) evolved and moderately massive ($\gtrsim 1.5 M_{\odot}$) stars can most readily survive many repeated stripping events, and 3) progressively dimmer flares -- observed in some repeating partial TDE candidates -- could be explained by the increase in the density of the star post-mass-loss. 
\end{abstract}


\section{Introduction}
\label{sec:intro}
Close encounters between supermassive black holes (SMBHs) and stars can give rise to the dynamical capture of a star by the SMBH 
following the tidal break-up of a binary (a.k.a.~the Hills mechanism; \citealt{hills88}). A star can also be partially destroyed by an SMBH if its pericenter distance $r_{\rm p}$ is comparable to its tidal radius $r_{\rm t} \equiv R_{\star} (M_{\bullet}/M_{\star})^{1/3}$ (where $M_{\bullet}$ is the mass of the SMBH, $M_{\star}$ and $R_{\star}$ are the mass and radius of the star, respectively; \citealt{hills75,rees88}). These two processes -- tidal capture and partial stellar destruction -- can act in tandem, such that the Hills-captured star is subsequently tidally stripped of mass, if the pericenter distance of the original binary is comparable to the tidal radius of (at least) one of the stars in the original binary. If the tidally stripped core of the star remains relatively structurally unchanged, it can continue to orbit the SMBH and lose mass on subsequent pericenter passages, powering recurrent accretion flares in a repeating partial tidal disruption event (rpTDE). This process can putatively explain a new class of observed nuclear transients that repeat on timescales of months to years, such as ASASSN-14ko \citep{payne21}, AT2018fyk \citep{wevers23}, eRASSt-J045650 \citep{liu24}, AT2020vdq \citep{somalwar23}, AT2022dbl \citep{lin24,hinkle24} and AT2021aeuk \citep{sun25}. For SMBHs with masses of the order $10^6M_{\odot}$ and solar-like stars, a period of $\sim 1$ year corresponds to a specific orbital energy of $\sim 20 GM_{\odot}/R_{\odot}$, implying that -- if there were no flares prior to the first detected outburst -- the Hills mechanism is required to place the star on its orbit about the SMBH, i.e., tides alone cannot bind the star sufficiently tightly to produce the observed months-to-years recurrence timescales \citep{cufari22, cufari23}.

One potential issue with the rpTDE model is that the cumulative effect of tidal heating over only a small number of encounters could lead to the inflation of the star, runaway mass loss, and its complete disruption~\citep{linial24,liu24}. This issue, and the more general question of the stability of a star undergoing multiple mass-loss events, has been analyzed recently through numerical hydrodynamical simulations, e.g.,~\cite{bandopadhyay24,liu25}. These works found that the survivability of a star largely depends on its internal structure, with more massive and centrally concentrated stars (generally chemically evolved and at a substantial age along the main sequence) 
being more likely to survive multiple encounters, with only a small amount ($\sim 1\%$ of the original mass of the star per encounter) of mass lost from the outer envelope per encounter and the central regions (hereafter the ``core'') relatively unaltered.

\citet{bandopadhyay24} also found that tidal dissipation did not result in the rapid destruction of the star, even when the energy dissipated numerically (viscously) -- which likely acts on timescales much shorter than the radiative damping time, but could be comparable to those related to nonlinear and turbulent dissipation \citep{zahn75, mcmillan87,kumar96,goodman98,weinberg12,weinberg19} -- was self-consistently maintained in the thermal energy content of the gas. This finding contrasts the expectation that results from the extrapolation of the linear theory, developed by \citet{fabian75, press77, lee86, mcmillan87} in the context of stellar binaries, which suggests that it should be more thermodynamically significant (see the discussion in \citealt{linial24}). However, the loss of mass from the star is a fundamentally nonlinear effect, and the accuracy of the linear formalism in describing this regime is questionable. Indeed, along these lines and in the context of hot Jupiters tidally excited by their host stars, \cite{faber05} found numerically (with Smoothed Particle Hydrodynamics; SPH) that the energy imparted to the planet is not enough to completely disrupt it, even in the limit where the planet loses mass through Roche lobe overflow (see also, e.g., \citealt{rasio96,ford99,ivanov04a,ivanov04b,ogilvie04}).

More generally, small changes to the internal structure of an object following such tidal interactions can have dramatic implications for subsequent encounters: because the tidal field scales with the inverse-cube of the distance between the star and the SMBH, small differences in the pericenter distance of the star result in substantial differences in the amount of mass lost via tides (e.g., \citealt{guillochon13, mainetti17, miles20, nixon21, coughlin22}). Correspondingly, a small reduction in the average density of the star following the removal of mass can greatly enhance its susceptibility to further tidal stripping, whereas an increase in its density would enable it to more readily resist the tidal influence of the SMBH. If -- depending on the original stellar properties -- one or the other of these outcomes is achieved following the first encounter, one could plausibly explain why some events, such as AT2018fyk \citep{wevers23} and eRASSt-J045650 \citep{liu24}, exhibit progressively weaker (in terms of luminosity) outbursts, while others, such as AT2020vdq \citep{somalwar23}, have brighter secondary flares. 

Completely mapping out this parameter space with high-resolution simulations, and thereby determining which stars in terms of mass, age, and (e.g.) metallicity are more or less susceptible to tidal stripping following the first encounter and/or undergo runaway tidal heating, is -- if not outright infeasible -- very computationally expensive and subject to questions regarding the accuracy and self-consistency of tidal dissipation. To surmount these issues, in this work we develop an intermediate/hybrid approach to understand the response of stars to mass loss. Specifically, we treat the star effectively as a Bonnor-Ebert sphere, such that the original progenitor is modeled as the combination of an outer ``confining medium'' that contains $x\%$ of the mass (where $x \lesssim 10$), and an inner ``core'' that contains the remaining $\left(100-x\right)\%$. The pressure of the surrounding medium is then reduced on the dynamical time of the core, thereby mimicking the mass loss, and the response of the core is reconstructed with linear perturbation theory. 
We model the response of a range of stars evolved with {\sc mesa}~\citep{paxton11,paxton13,paxton15,paxton18} to varying degrees of mass loss, and analyze the differences in stellar structure that could give rise to various qualitative differences in the lightcurves of rpTDEs.

In Sections~\ref{sec:fluid-eqns}-\ref{sec:energy} we outline our framework for studying the response of a star to mass loss, which consists of the superposition of a background state and perturbations. The background state is obtained by modeling the mass loss as occurring through a sequence of quasi-steady states, such that as the pressure at the outer surface is reduced (i.e., as the mass is removed) the star always retains hydrostatic balance. The core expands through this sequence of quasi-steady equilibria, and relaxes to an asymptotic, time-independent background state. Perturbations are excited as a consequence of the velocity associated with the outward expansion of the surface, which we quantitatively analyze with the eigenmodes of the asymptotic background state, i.e., we solve for the secular expansion of the stellar surface and the subsequent oscillations associated with the kinetic energy of the gas. We apply this methodology to {\sc mesa}-generated \citep{paxton11, paxton13, paxton15, paxton18}, stars in Section~\ref{sec:applications}, and compare the results to -- and find excellent agreement with -- 1D hydrodynamical simulations conducted with the finite-volume code {\sc flash} \citep{fryxell00}.

In Section \ref{sec:phantom} we compare the predictions of our model to smoothed-particle hydrodynamics simulations of the partial tidal disruption of stars by SMBHs. Despite the obvious qualitative differences between our simplified physical picture -- a spherically symmetric object undergoing pressure de-confinement -- and the nonlinear tidal interaction between a star and an SMBH, the following predictions of the model, the details and implications of which are discussed in Section \ref{sec:implications}, are upheld: 1) higher-mass ($M_{\star} \gtrsim 0.7 M_{\odot}$) stars (lower-mass stars) undergo an increase (a decrease) in volume-averaged density following the removal of a small amount of mass, rendering them less susceptible (more susceptible) to subsequent mass loss; 
2) the energy of the surviving core is primarily determined by the difference between the binding energy of the star prior to mass loss and the initial binding energy of the core, and the $p-\mathrm{d}V$ work done by the stellar surface as it expands following the mass loss leads to a small negative correction to the total energy content of the surviving core; 
3) the kinetic energy associated with the oscillatory modes of the star, which is ultimately dissipated as heat, is substantially smaller than the change in the internal energy, the gravitational energy, and the sum of the two, and is therefore a) effectively irrelevant as concerns the post-mass-loss stellar structure, and b) suggestive of the fact that runaway tidal heating is not an issue over many (tens to hundreds, depending on the amount of mass lost) repeated encounters. We also find that the energy in rotation, which is not included in our model (but is in the numerical simulations), is second in importance to the difference in binding energies resulting from the removal of mass from the outer layers of the star, i.e., it is larger in magnitude than the $p-\mathrm{d} V$ work done by the expansion of the surface. The rotational energy therefore contributes non-negligibly to the final energy of the surviving core, but only at the $\sim 10\%$ level compared to the difference in binding energy of the core pre-mass-loss and that of the initial star.

We summarize our results, as well as comment on the applicability of this model to other astrophysical systems, in Section~\ref{sec:summary}.

\section{Lagrangian mass loss model}
\label{sec:lagrangian-model}
We divide the original star into two regions: the core, which extends to an original radius $R_{\rm 0, c}$ in the star and has mass $M_{\rm c}$, and the envelope, which extends from $R_{\rm 0, c}$ to $R_{\star}$ and contains mass $M_{\star}-M_{\rm c}$. The effects of the mass loss can then be modeled by considering the response of the core to a reduction in the pressure of the surrounding envelope, i.e., by reducing the pressure of the fluid shell with Lagrangian radius $R_{\rm 0, c}$ over some timescale that is comparable to the dynamical time of the star. 

If the relative change in the pressure at $R_{\rm 0, c}$ were small, then the response of the core could be modeled self-consistently through linear perturbation theory, with the background state of the star being the initial, pressure-confined core. The drop in pressure then acts as a driving term that stimulates the expansion of the core and subsequent oscillations about a new, average radius. However, here we are interested in the limit that the surrounding pressure goes to zero, implying that the relative change in pressure is eventually not small. 

We show that this issue can be overcome by treating the background state of the star not as the initial hydrostatic core, but as a sequence of quasi-steady states, each of which exactly matches the pressure of the surrounding medium at any instant in time. The system approaches a final, time-averaged configuration that has zero pressure and density at its surface. The core also possesses oscillatory motion, and we show that this motion can be related to the time dependence of the background state. We go on to predict the complete temporal evolution of the new stellar surface, from its initial expansion to subsequent oscillations, the time-averaged density of the surviving core relative to the initial stellar density, and the energy contained in the final background state and the oscillatory modes. 

\subsection{Lagrangian fluid equations}
\label{sec:fluid-eqns}
If the gas obeys an adiabatic equation of state with adiabatic index $\Gamma$, the spherically symmetric continuity and entropy equations are, in Lagrangian form, 
\begin{equation}
\begin{split}
    &\rho(r_0,t) = \rho_0(r_0)\left(\frac{\partial r}{\partial r_0}\right)^{-1}\left(\frac{r(r_0,t)}{r_0}\right)^{-2}, \\
   & p(r_0,t) = p_0(r_0)\left(\frac{r(r_0,t)}{r_0}\right)^{-2\Gamma}\left(\frac{\partial r}{\partial r_0}\right)^{-\Gamma}, \label{content}
\end{split}
\end{equation}
where $\rho_0$ and $p_0$ are the initial density and pressure profiles of the star and $r_0$ is the initial Lagrangian radius of a fluid element. Then the radial momentum equation for the Lagrangian displacement $r(r_0,t)$ is
\begin{multline}
    \frac{\partial^2 r}{\partial t^2}+\frac{1}{\rho_0(r_0)}\left(\frac{r}{r_0}\right)^2\frac{\partial}{\partial r_0}\left[p_0(r_0)\left(\frac{r}{r_0}\right)^{-2\Gamma}\left(\frac{\partial r}{\partial r_0}\right)^{-\Gamma}\right] \\
    = -\frac{GM_0(r_0)}{r^2}, \label{rmom}
\end{multline}
where $M_0(r_0)$ is the initial mass profile of the star. 

We let our system be described by the above set of equations out to some initial Lagrangian radius $R_{\rm 0, c}$, the solution interior to this radius being the core. Outside of this radius, which we denote the envelope, we do not model the fluid dynamics, but rather allow the pressure of this region to decline with time. 
Denoting the pressure at this radius by $p_{\rm c}(t)$, the boundary condition on our fluid at the surface of the core is
\begin{equation}
p_{\rm c}(t) = p_{\rm 0}(R_{\rm 0, c})\left(\frac{r(R_{\rm 0, c}, t)}{R_{\rm 0, c}}\right)^{-2\Gamma}\left(\frac{\partial r}{\partial r_0}\bigg{|}_{R_{\rm 0, c}}\right)^{-\Gamma}.  \label{pbc}
\end{equation}
Equations \eqref{content} -- \eqref{pbc} are general. If the pressure of the surrounding medium drops to zero, then Equation \eqref{pbc} illustrates that -- assuming the final radius of the core remains finite -- the derivative $\partial r/\partial r_0$ will diverge at that radius, i.e., the new radius of the core coincides with zero density. 

\subsection{Background state}
\label{sec:background}
The background state, which we denote as $r_{\rm b}(r_0,t)$, is one that proceeds through a sequence of quasi-steady states that satisfies the exact pressure boundary condition \eqref{pbc} at any time. We further non-dimensionalize the fluid variables by defining
\begin{equation}
\begin{split}
   & \xi = \frac{r}{R_{\rm 0, c}}, \,\,\, \xi_0 = \frac{r_0}{R_{\rm 0,c}}, \,\,\, \tau = \frac{\sqrt{GM_{\rm c}}}{R_{\rm 0, c}^{3/2}}t, \,\,\, M_0 = M_{\rm c}m_0(\xi_0), \\
  &  \,\,\,  \rho_0 = \frac{M_{\rm c}}{4\pi R_{\rm 0, c}^3}g_0(\xi_0), \,\,\, p_0 = \frac{GM_{\rm c}^2}{4\pi R_{\rm 0,c}^4}h_0(\xi_0),
\end{split}
\end{equation}
where $R_{\rm 0, c}$ is the initial Lagrangian radius of the core (i.e., before the pressure is dropped) and $M_{\rm c}$ is the mass enclosed at that radius. The background state, $\xi_{\rm b} = r_{\rm b}(\xi_0,t)/R_{\rm 0, c}$, thus satisfies
\begin{equation}
    \frac{1}{g_0}\left(\frac{\xi_{\rm b}}{\xi_0}\right)^2\frac{\partial}{\partial \xi_0}\left[h_0\left(\frac{\xi_{\rm b}}{\xi_0}\right)^{-2\Gamma}\left(\frac{\partial \xi_{\rm b}}{\partial \xi_0}\right)^{-\Gamma}\right]=-\frac{m_0}{\xi_{\rm b}^2}, \label{xibeq}
\end{equation}
alongside the surface boundary condition
\begin{equation}
    \left(\frac{\partial\xi_{\rm b}}{\partial \xi_0}\bigg{|}_{\xi_0 = 1}\right)^{-\Gamma}\xi_{\rm b}(1)^{-2\Gamma} = \frac{p_{\rm c}(t)}{p_0(R_{\rm 0,c})} \equiv h_{\rm c}(t), \label{xibbc}
\end{equation}
where $h_{\rm c}$ is the pressure at the surface relative to the initial pressure at that location. The background state therefore depends only on the initial core radius (or, equivalently, on the mass lost) and only implicitly on time through the pressure at the core radius. 
Equation \eqref{xibeq} can be integrated from $\xi_0 = 1$ for an arbitrary value of $\xi_{\rm b}(1)$, with $\partial\xi_{\rm b}/\partial \xi_0(1)$ from Equation \eqref{xibbc}. The resulting solution will generally terminate before reaching the origin or will not satisfy $\xi_{\rm b}(0) = 0$, except for a special value of $\xi_{\rm b}(1)$; we shoot from the surface and iterate on $\xi_{\rm b}(1)$ until the solution satisfies $\xi_{\rm b}(0) \simeq 0$, which then determines the solution. 

\begin{figure*}
    \includegraphics[width=0.495\textwidth]{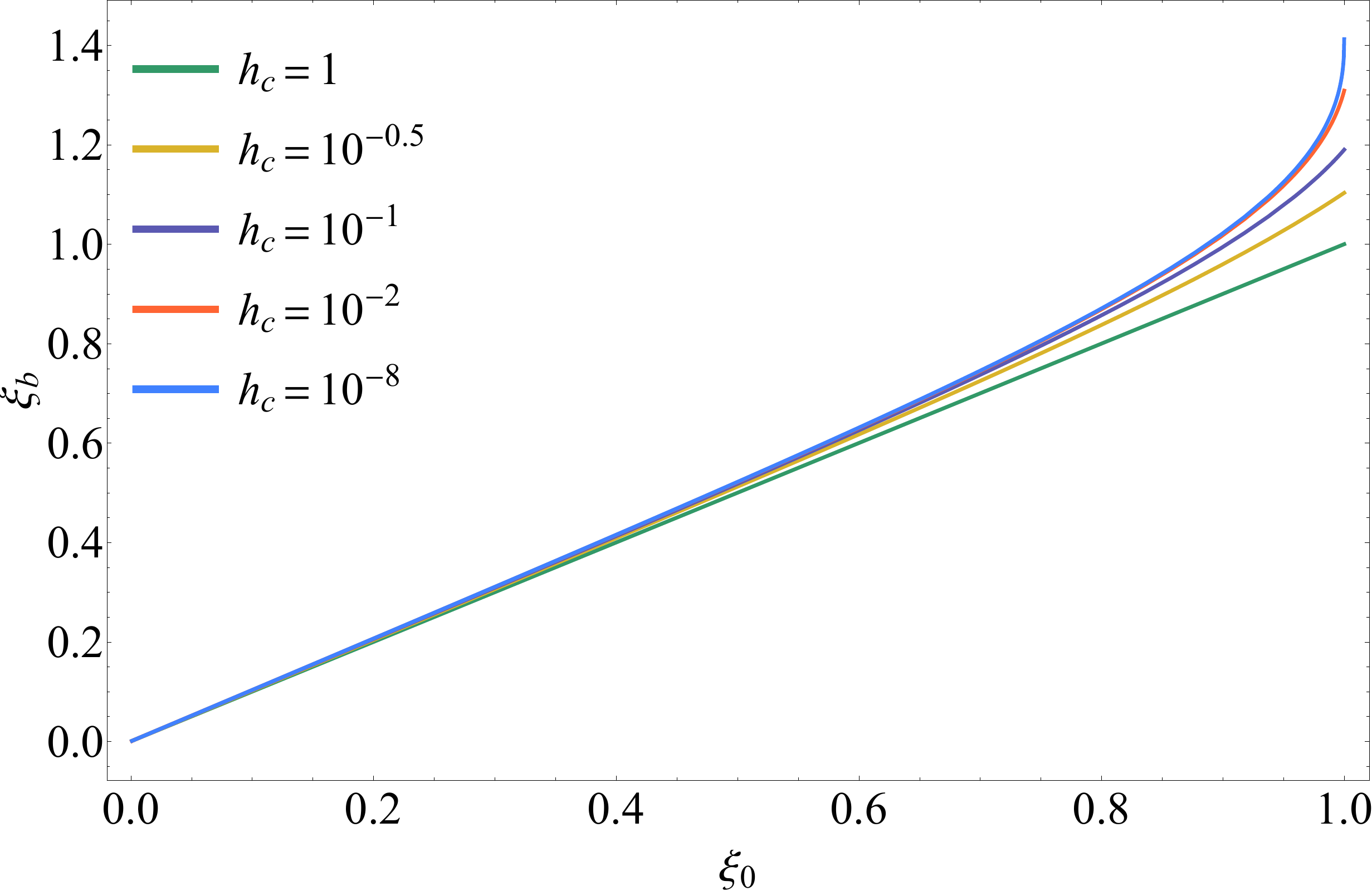}
    \includegraphics[width=0.495\textwidth]{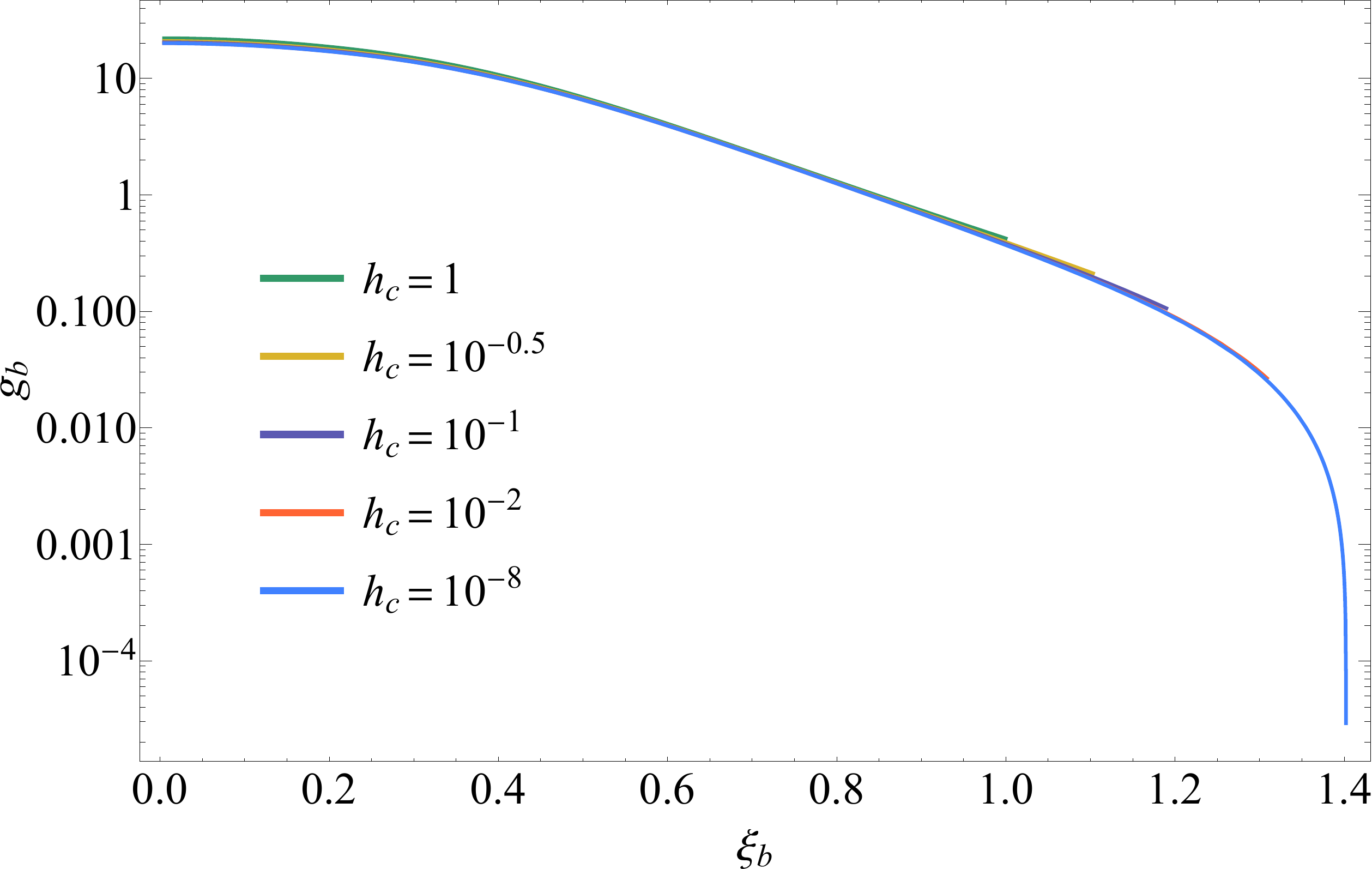}
    \caption{Left: the background state of the star, $\xi_{\rm b}$, as a function of the initial Lagrangian radius in the core, $\xi_{0}$, for the relative surface pressures $h_{\rm c}$ shown in the legend and a mass loss of $10\%$. As the pressure declines the system approaches a final equilibrium, and $\partial\xi_{\rm b}/\partial\xi_0(\xi_0 = 1)\rightarrow \infty$. Right: the dimensionless density of the background state, $g_{\rm b}$, as a function of the instantaneous/current positions of the fluid elements $\xi_{\rm b}$, at the same surfaces pressures as the left panel. Here $g_{\rm b}$ is proportional to the initial density but scaled by $\left(\xi_{\rm b}/\xi_0\right)^{-2}\left(\partial\xi_{\rm b}/\partial \xi_0\right)^{-1}$, i.e., it is the density that is derived from the initial mass profile of the star with the radii of fluid elements stretched according to $\xi_{\rm b}(\xi_0)$. Written in terms of $\xi_{\rm b}$, it is clear that the surface expands outward as the pressure drops.}
    \label{fig:xib_of_xi0}
\end{figure*}

Figure \ref{fig:xib_of_xi0} gives an example of the solution for $\xi_{\rm b}(\xi_0, h_{\rm c})$ (left panel) for a $1M_{\odot}$ star at the zero-age main sequence (ZAMS), constructed with {\sc mesa}, from which 10\% of the mass was removed. In this case the initial core radius (which encloses 10\% of the mass of the star) is $R_{\rm 0, c} \simeq 0.494 R_{\odot}$, compared to the initial stellar radius of $R_{\star} \simeq 0.888 R_{\odot}$, which highlights the fact that such a star is somewhat centrally concentrated (i.e., $R_{\rm 0, c}/R_{\star} \simeq 0.56$, even though we are only removing 10\% of the mass). The values of the pressure at the surface are shown in the legend, and for $h_{\rm c} \lesssim 10^{-3}$, the solution approaches a final equilibrium for which the density (and pressure) is $\sim$ zero at the surface. The final dimensionless radius of the star is, in this case, $\xi_{\rm b} \simeq 1.41$, or a physical radius of $R_{\rm c} \simeq 1.41 R_{\rm 0, c} \simeq 0.70 R_{\odot}$, which is less than the initial stellar radius of $\sim 0.89 R_{\odot}$. 

The right panel of Figure \ref{fig:xib_of_xi0} gives the density of the star as a function of $\xi_{\rm b}$, rather than $\xi_0$, to highlight the fact that the radius of the star moves out (this is also, obviously, apparent from the left panel), where
\begin{equation}
    g_{\rm b} = \frac{1}{\xi_{\rm b}^2}\left(\frac{\partial \xi_{\rm b}}{\partial \xi_0}\right)^{-1}\frac{dm_0}{d\xi_0} = \left(\frac{\xi_{\rm b}}{\xi_0}\right)^{-2}\left(\frac{\partial \xi_{\rm b}}{\partial \xi_0}\right)^{-1}g_0 \label{gb}
\end{equation}
is the density profile in terms of $\xi_{\rm b}$. In this expression we write $\xi_0(\xi_{\rm b})$, and thereby invert the solution shown in the left-hand panel, to plot the density in terms of the current radial coordinate in the star. By the point where the external pressure drops to $h_{\rm c} = 10^{-8}$, the density at the core radius has fallen to $\lesssim few\times 10^{-5}$, and the solution does not evolve significantly once $h_{\rm c} \lesssim 10^{-3}$.

\begin{figure}
    \includegraphics[width=0.495\textwidth]{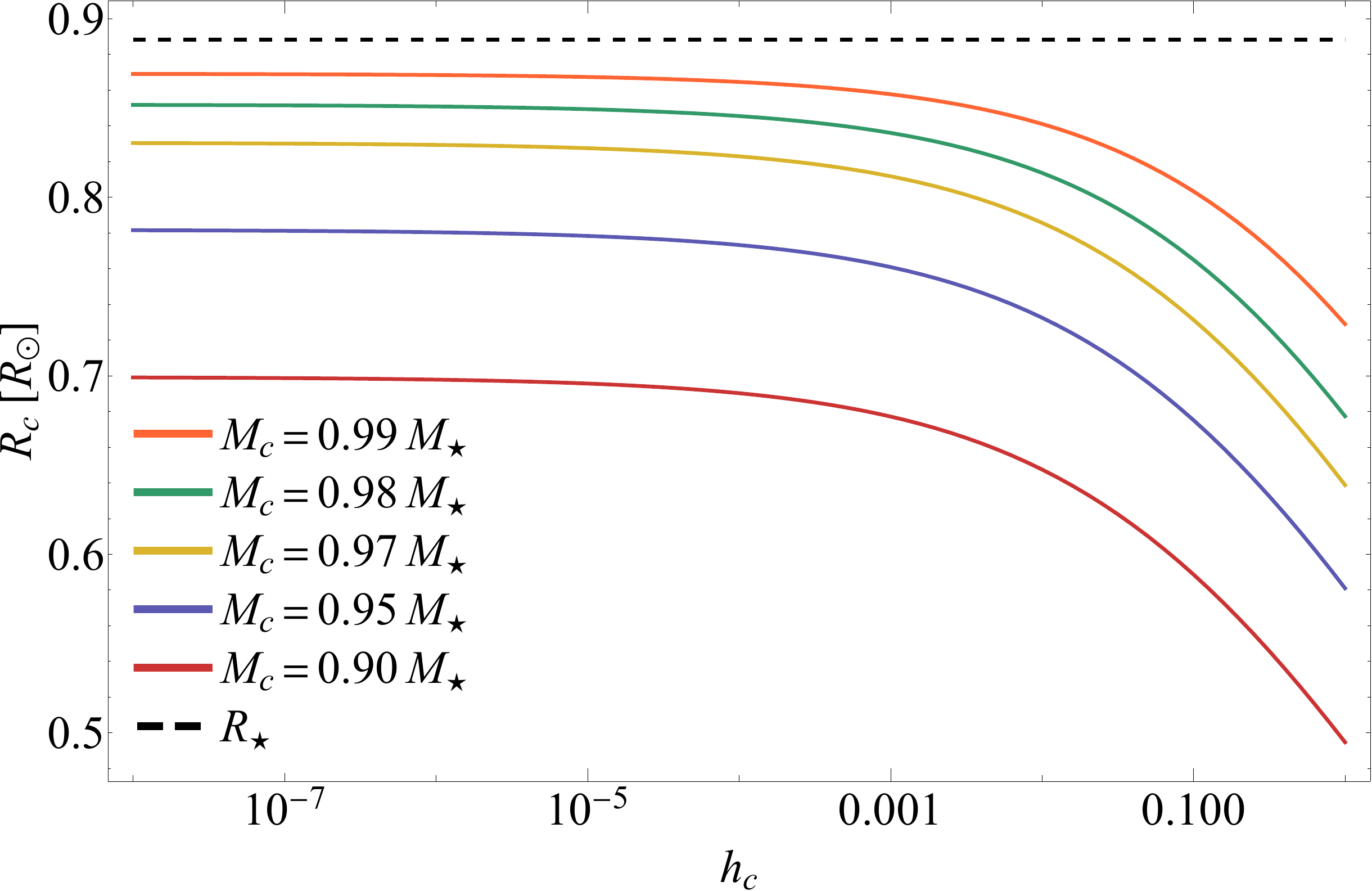}
      \caption{The core radius as a function of surface pressure, showing that the radius expands as the pressure drops (i.e., time goes from right to left on the horizontal axis) and approaches a limiting value as $h_{\rm c} \rightarrow 0$. The different curves are appropriate to the core masses in the legend, while the black-dashed line is the initial radius of the star. For this star, the final core radius is therefore always less than the initial stellar radius. }
    \label{fig:Rc_of_hc}
\end{figure}

Figure \ref{fig:Rc_of_hc} gives the core radius as a function of the surface pressure for the same, $1 M_{\odot}$ star and for the core masses shown in the legend. In each case the core radius starts at a value that is significantly less than the original stellar radius (shown by the black-dashed line), but increases and asymptotes to a constant value as the surface pressure drops to zero. 

\subsection{Perturbations}
\label{sec:perturbations}
The background state depends on the pressure at the core radius, which itself depends on time (in some way that must be externally imposed in this model). The fluid shells therefore have a finite velocity when the pressure drops over a finite timescale, which is not self-consistently included in our definition of the background state. We can analyze the effects of the core velocity by writing the Lagrangian position of a fluid element as
\begin{equation}
    \xi = \xi_{\rm b}(\xi_0, p_{\rm c}(\tau))+\xi_1(\xi_0,\tau), \label{xiexp}
\end{equation}
where $\xi_1$ is assumed to be a small correction that is induced by the explicit time dependence of the expanding core. We can insert Equation \eqref{xiexp} into the momentum equation, expand the solution about the background state, and retain first-order terms in $\xi_1$; the result is
\begin{multline}
    \frac{\partial^2\xi_1}{\partial \tau^2}-\frac{\Gamma}{g_0}\left(\frac{\xi_{\rm b}}{\xi_0}\right)^2\frac{\partial}{\partial \xi_0}\bigg[h_0(\xi_0)\left(\frac{\xi_{\rm b}}{\xi_0}\right)^{-2\Gamma}\left(\frac{\partial\xi_{\rm b}}{\partial \xi_0}\right)^{-\Gamma} \\ 
    \times\left(\frac{2\xi_1}{\xi_{\rm b}}+\left(\frac{\partial \xi_{\rm b}}{\partial \xi_0}\right)^{-1}\frac{\partial\xi_1}{\partial \xi_0}\right)\bigg]
    -\frac{4 m_0(\xi_0)}{\xi_{\rm b}^3}\xi_1 = -\frac{\partial^2\xi_{\rm b}}{\partial \tau^2}. \label{xi1eq}
\end{multline}
We can take the Laplace transform of the preceding equation, where
\begin{equation}
    \tilde{\xi}_1 = \int_0^{\infty}\xi_1(\tau)e^{-\sigma\tau}d\tau
\end{equation}
is the Laplace transform of $\xi_1$, and
\begin{equation}
    \int_0^{\infty}\frac{\partial^2\xi_1}{\partial \tau^2}e^{-\sigma\tau}d\tau = \sigma^2\tilde{\xi}_1-\frac{\partial\xi_1}{\partial \tau}\bigg{|}_{\tau = 0}-\sigma\xi_1(\tau=0).
\end{equation}
At $\tau = 0$ we require $\xi = \xi_0$, and since $\xi_{\rm b}(\tau =0) = \xi_0$, we have $\xi_1(\tau = 0) = 0$. The star must also be in hydrostatic balance immediately prior to mass loss, such that $\partial\xi/\partial\tau(\tau = 0) = 0$. To maintain this condition, we therefore also demand that
\begin{equation}
    \frac{\partial\xi_1}{\partial \tau}\bigg{|}_{\tau = 0} = -\frac{\partial\xi_{\rm b}}{\partial \tau}\bigg{|}_{\tau = 0} = -\frac{\partial h_{\rm c}}{\partial \tau}\bigg{|}_{\tau = 0}\frac{\partial \xi_{\rm b}}{\partial h_{\rm c}}\bigg{|}_{h_{\rm c} = 1}.
\end{equation}

There are thus two terms that are responsible for driving the perturbations, the first of which is the acceleration of the surface, $\partial^2\xi_{\rm b}/\partial\tau^2$, which can also be written in terms of the temporal derivatives of the pressure. Given that the background state approaches a new equilibrium as the surface pressure declines, this contribution will tend toward zero in the limit of large-$\tau$, such that the Laplace transform yields terms that -- upon being inverse-transformed -- asymptotically decay. The same is not true for the initial velocity, which contributes a constant (i.e., $\sigma$-independent) influence on the state to asymptotically late times. Furthermore, since the background state approaches a new equilibrium as the pressure nears zero, we can approximate $\xi_{\rm b}$ on the left-hand side of Equation \eqref{xi1eq} as this asymptotic state, i.e., $\xi_{\rm b}$ becomes time-independent. We can then define $\partial h_{\rm c}/\partial \tau(\tau = 0) \equiv \omega$, where we expect $\omega \sim 1$ for the case of a TDE where the mass is removed on $\sim$ the dynamical time of the star, and Equation \eqref{xi1eq} becomes
\begin{equation}
    \sigma^2\tilde{\xi}_1-\frac{\Gamma}{g_{\rm b}}\frac{\partial}{\partial \xi_{\rm b}}\left[\frac{h_{\rm b}}{\xi_{\rm b}^2}\frac{\partial}{\partial \xi_{\rm b}}\left[\xi_{\rm b}^2\tilde{\xi}_1\right]\right]-\frac{4 m_{\rm b}}{\xi_{\rm b}^3}\tilde{\xi}_1 = -\omega\frac{\partial \xi_{\rm b}}{\partial h_{\rm c}}\bigg{|}_{h_{\rm c} = 1}. \label{xi1teq}
\end{equation}
Here $g_{\rm b}$ is given by Equation \eqref{gb}, $m_{\rm b}(\xi_{\rm b}) = m_0(\xi_0(\xi_{\rm b}))$, and 
\begin{equation}
    h_{\rm b}(\xi_{\rm b}) \equiv h_0(\xi_{\rm b})\left(\frac{\xi_{\rm b}}{\xi_0}\right)^{-2\Gamma}\left(\frac{\partial\xi_{\rm b}}{\partial \xi_0}\right)^{-\Gamma}.
\end{equation}
Note that, to avoid introducing more variables, we are using $\xi_{\rm b}$ to refer to the asymptotic background state as the pressure drops to zero in Equation \eqref{xi1teq}. The function $\partial\xi_{\rm b}/\partial h_{\rm c}(h_{\rm c} = 1)$ depends only on the star and the core radius (mass lost), and hence it is only $\omega$, the temporal derivative of the pressure at $\tau = 0$ -- rather than the full functional form of the pressure with respect to time, which in our case is arbitrary -- that drives the perturbations in this approximation. 

The spatial operator in Equation \eqref{xi1teq} is in Sturm-Liouville form, and we can expand the solution for the perturbations in terms of the eigenmodes of that spatial operator, i.e., the eigenmodes $\xi_{\rm n}(\xi_{\rm b})$ satisfy
\begin{equation}
    \frac{\Gamma}{g_{\rm b}}\frac{\partial}{\partial \xi_{\rm b}}\left[\frac{h_{\rm b}}{\xi_{\rm b}^2}\frac{\partial}{\partial \xi_{\rm b}}\left[\xi_{\rm b}^2\xi_{\rm n}\right]\right]+\frac{4 m_{\rm b}}{\xi_{\rm b}^3}\xi_{\rm n} = -\sigma_{\rm n}^2\xi_{\rm n},
\end{equation}
alongside the boundary conditions
\begin{equation}
    \xi_{\rm n}(\xi_{\rm b}(1)) = 1, \,\,\, \frac{\partial \xi_{\rm n}}{\partial \xi_{\rm b}}\bigg{|}_{\xi_{\rm b}(1)} = \frac{\xi_{\rm b}(1)^2}{\Gamma}\left(\sigma_{\rm n}^2+\frac{4-2\Gamma}{\xi_{\rm b}(1)^3}\right).
\end{equation}
The eigenvalues $\sigma_{\rm n}$ are determined by requiring that $\xi_{\rm n}(0) = 0$, and can be recovered numerically with the same shooting method used to determine the background state (see the previous subsection). The eigenmodes are then orthogonal with respect to the weight $g_{\rm b}\xi_{\rm b}^2$, and hence the time-dependent response of the star is
\begin{equation}
    \xi_1(\xi_{\rm b}, \tau) = \omega\sum_{n} c_{\rm n}\frac{\sin\left(\sigma_{\rm n}\tau\right)}{\sigma_{\rm n}}\xi_{\rm n}(\xi_{\rm b}), \label{xi1sum}
\end{equation}
where $c_{\rm n}$ is the following (normalized) overlap integral:
\begin{equation}
    c_{\rm n} = \left(\int_0^{\xi_{\rm b}(1)}\xi_{\rm n}^2dm_{\rm b}\right)^{-1}\int_0^{\xi_{\rm b}(1)}\xi_{\rm n}\frac{\partial\xi_{\rm b}}{\partial h_{\rm c}}\bigg{|}_{h_{\rm c} = 1}dm_{\rm b}, \label{coeffs}
\end{equation}
where $dm_{\rm b} = g_{\rm b}\xi_{\rm b}^2d\xi_{\rm b}$.

\begin{figure*}
    \includegraphics[width=0.485\textwidth]{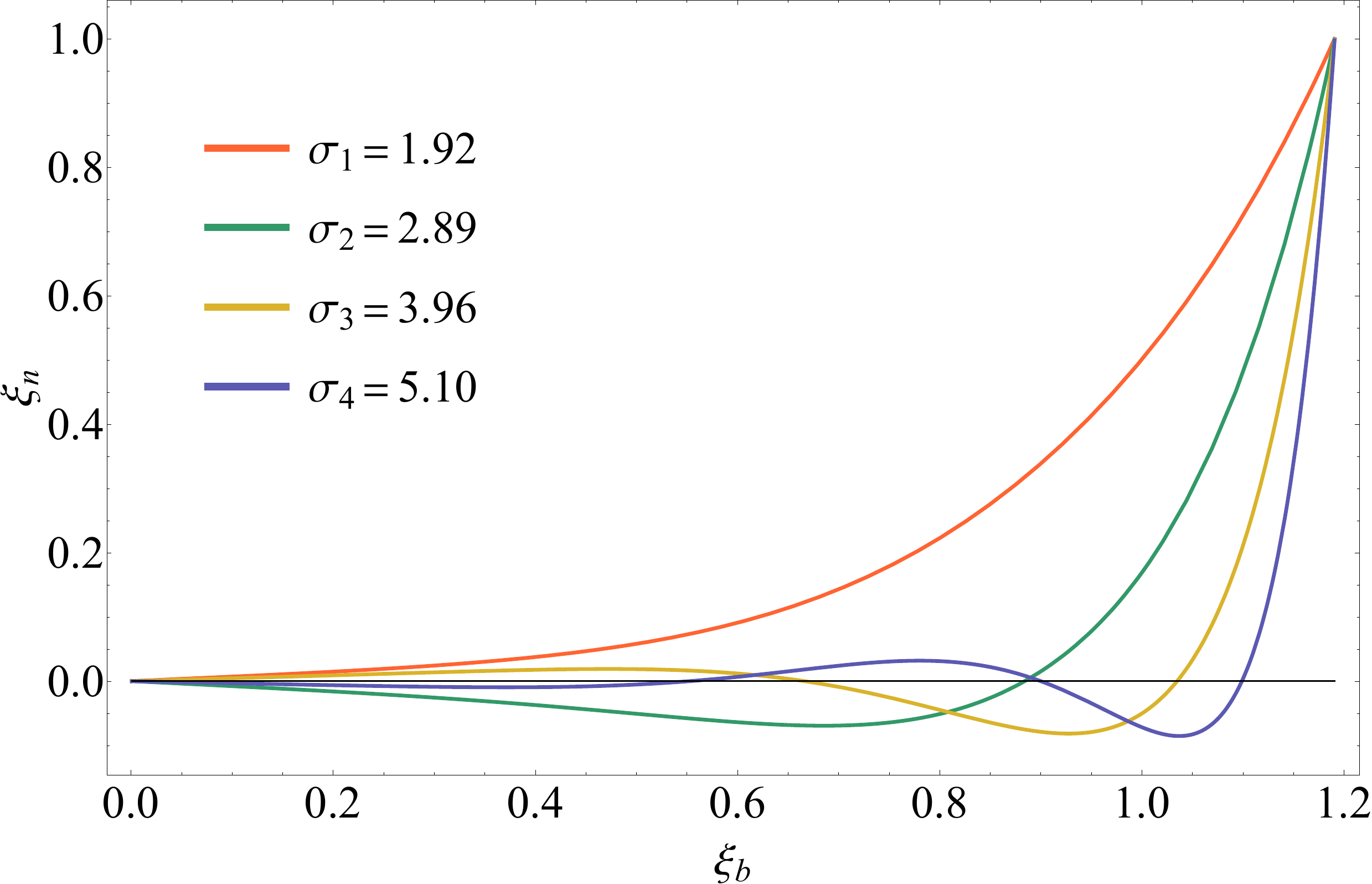}
     \includegraphics[width=0.505\textwidth]{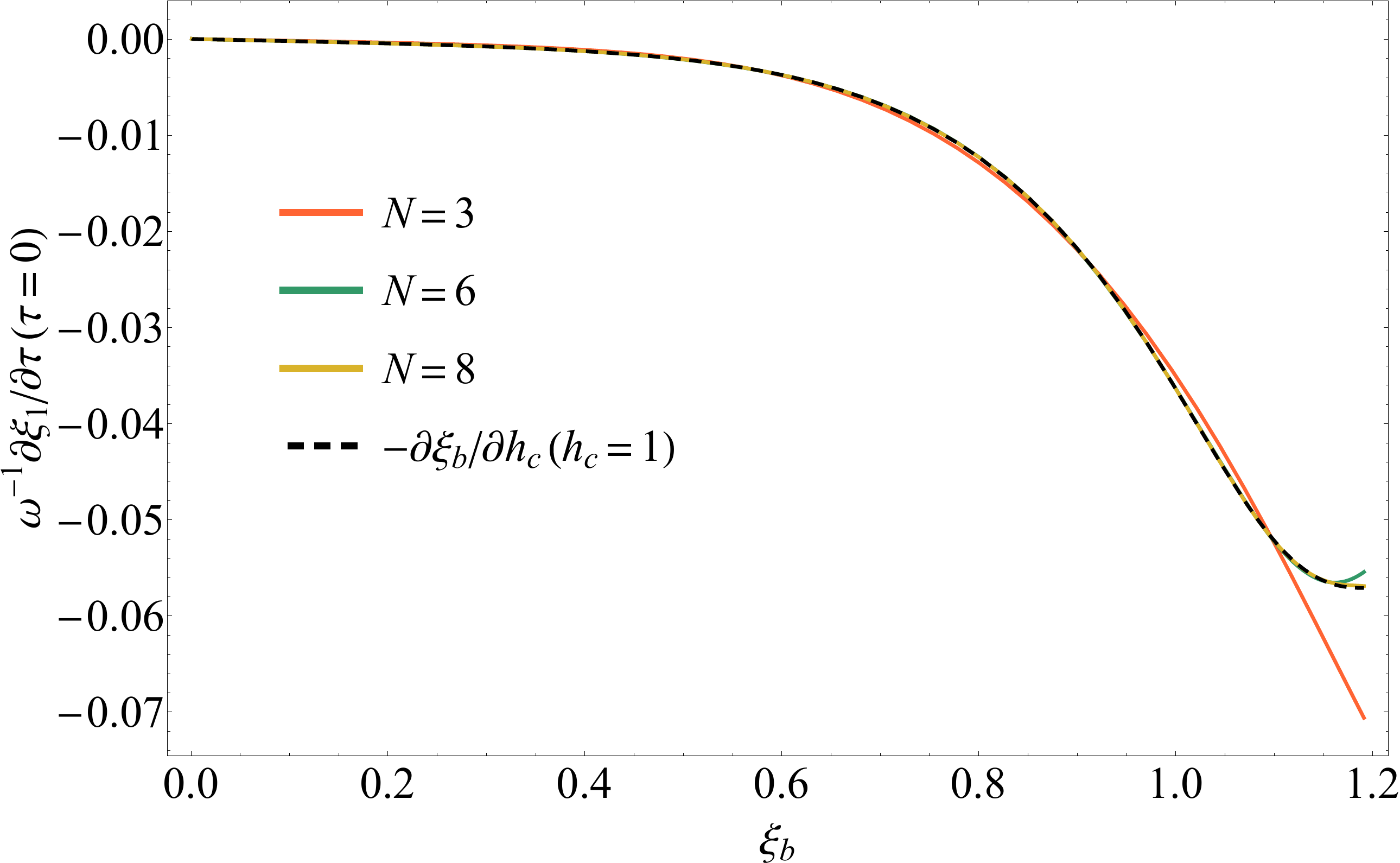}
    \caption{Left: the first four eigenmodes describing the background state of the star for a $1M_{\odot}$, ZAMS star with $1\%$ mass loss, after the pressure at the surface is reduced by a factor of $10^{-6}$. The eigenvalues appropriate to the modes are shown in the legend. Right: the initial velocity profile of the core normalized by $\omega$, shown by the black-dashed curve. The orange, green, and yellow curves illustrate the reconstruction of the initial velocity profile with $3$, $6$, and $8$ modes, respectively, which demonstrates that only $\lesssim 10$ eigenmodes are required to accurately analyze the behavior of the star. }
    \label{fig:modes}
\end{figure*}

The left panel of Figure \ref{fig:modes} shows, as an example, the first four modes (alongside the eigenvalues in the legend) for a $1M_{\odot}$ star at ZAMS with 1\% mass loss (i.e., the core contains 99\% of the mass). Here the pressure at the surface of the background state of the star has fallen to $10^{-6}$ of its original value, such that it has effectively reached a new equilibrium (i.e., further reducing the pressure effectively leaves the background state unchanged). As expected, each successive mode has one more zero crossing, and the power of the modes becomes increasingly confined to the surface as the mode number increases. 

The number of modes required to construct the solution for the velocity to high accuracy is on the order of $\sim 10$, and is considerably fewer for the time-dependent response of the surface, owing to the fact that the coefficients are multiplied by an additional factor of $1/\sigma_{\rm n}$ when calculating the displacement (i.e., Equation \ref{coeffs}). To see this explicitly, the right panel of Figure \ref{fig:modes} shows the reconstruction of the initial velocity profile of the background state (shown by the black, dashed curve) alongside the mode reconstruction, where $N$ indicates the maximum number of modes used in the sum in Equation \eqref{xi1sum}. By 6 modes the two agree extremely well, with only a slight disagreement near the surface, and the two are effectively indistinguishable with $N = 8$. This figure simultaneously validates the approach and highlights the accuracy with which the eigenmodes are determined and their corresponding orthogonality.

\subsection{Total solutions}
We can approximate the total solution for the fluid by superimposing the background state and the perturbations derived in the previous section, i.e., 
\begin{equation}
    \xi(\xi_0, \tau) = \xi_{\rm b}(\xi_0,\tau)+\xi_1(\xi_0,\tau),
\end{equation}
where the dependence of $\xi_1$ on $\xi_0$ is through the relationship between the asymptotic state obtained by $\xi_{\rm b}$ (see the preceding subsection). The time-dependent radius of the core is then given by this expression evaluated at $\xi_0 = 1$, such that if we define the dimensionless background core radius as $\xi_{\rm c}(\tau) = \xi_{\rm b}(\xi_0 = 1,\tau)$, we have
\begin{equation}
    R_\mathrm{c} (\tau) = R_\mathrm{0,c} \left(\xi_{\rm c}(\tau) + \xi_1(1,\tau)\right).
\end{equation}
Even though this ignores effects that are related to the acceleration of the background state, both in terms of driving the oscillations and modifying the mode frequencies at early times, we still expect this approximation to provide an accurate representation of both the initial expansion of the surface and the subsequent oscillations about the asymptotic background state; following a discussion of the energy of the star, we show this explicitly in Section \ref{sec:applications}.

\subsection{Energy Imparted to the Star}
\label{sec:energy}
The time rate of change of the total energy of the core, $E_{\rm c}$, is decremented by the $p-dV$ work done by the expansion of the core as the pressure at the surface drops:
\begin{equation}
    \frac{\partial E_{\rm c}}{\partial t} = -4\pi p(R_{\rm c})R_{\rm c}^2\frac{\partial R_{\rm c}}{\partial t}. \label{dEtotdt}
\end{equation}
The core energy $E_{\rm c} = \mathcal{T}+\mathcal{V}+\mathcal{U}$ is the sum of the kinetic, $\mathcal{T}$, gravitational, $\mathcal{V}$, and thermal energies, $\mathcal{U}$, which are individually
\begin{equation}
    \mathcal{T} = 4\pi\int_0^{R_{\rm 0, c}}\frac{1}{2}v^2\rho_0(r_0)r_0^2dr_0, 
\end{equation}
\begin{equation}
    \mathcal{V} = -4\pi\int_0^{R_{\rm 0, c}}\frac{GM_0(r_0)}{r}\rho_0 r_0^2 dr_0,
\end{equation}
\begin{equation}
    \mathcal{U} = 4\pi \int_0^{R_{\rm 0, c}}\frac{1}{\Gamma-1}\frac{p}{\rho}\rho_0(r_0) r_0^2 dr_0.
\end{equation}
The time-dependent energy of the core is then obtained by integrating Equation \eqref{dEtotdt} with respect to time, such that the final energy of the core (i.e., after the surface pressure drops to zero at asymptotically late times) is
\begin{equation}
    E_{\rm c, final} = E_{\rm c, 0}-4\pi\int_0^{\infty}p(R_{\rm c})R_{\rm c}^2\frac{\partial R_{\rm c}}{\partial \tau}\,d\tau, \label{DeltaE}
\end{equation}
where $E_{\rm c, 0}$ is the energy of the core prior to the mass loss. 

We can use our expression for the time-dependent radius of the core to evaluate the contribution to this integral from the background state. Furthermore, since the background state depends only on $h_{\rm c}$, the zeroth-order change in the energy does not depend on the temporal evolution of the surface pressure. Specifically, we find
\begin{equation}
\begin{split}
    E_{\rm c, final} &= E_{\rm c, 0}+\frac{4\pi}{3}p_{0, \rm c}R_{0,\rm c}^3\left(1-\int_0^{1}\xi_{\rm c}(h_{\rm c})^3\,dh_{\rm c}\right) \\
    &\equiv E_{\rm c, 0} + \Delta E_{\rm p-dV}, \label{Ecfin}
\end{split}
\end{equation}
where $p_{0, \rm c}$ is the pressure at the initial radius of the core; we also defined the last term in the first equality as $\Delta E_{\rm p-dV}$, which is the $p-\mathrm{d}V$ work done by the surface.
Since $\xi_{\rm c} \ge 1$ (i.e., the core radius expands as the pressure drops), it follows that $E_{\rm c, final} < E_{\rm c, 0}$, and hence the expansion of the surface  results in the core becoming {more bound} to itself. 

However, this does not necessarily imply that the final core is more bound than the initial (core+envelope) star. In particular, the specific energy of the gas is generally negative near the stellar surface, such that by removing the outer stellar layers the core is less gravitationally bound than the original star. The total change in the energy relative to the initial energy of the star is then
\begin{equation}
    \Delta E = E_{\rm c, final}-E_{\star},
\end{equation}
such that if $\Delta E > 0$, the core is less gravitationally bound compared to the original star. We emphasize, however, that this change is due entirely to the removal of mass, and is independent of any additional energy imparted to the system via oscillatory modes. 

These oscillatory/higher-order terms can, at late times when the background state is no longer evolving temporally, be evaluated with the integral expressions for the individual components of the energy. Specifically, the kinetic energy becomes
\begin{equation}
    \mathcal{T}(\tau \rightarrow \infty) = \frac{1}{2}\frac{GM_{\rm c}^2}{R_{\rm 0, c}}\int_0^{1}\left(\frac{\partial\xi_1}{\partial \tau}\right)^2g_0 \xi_0^2d\xi_0,
\end{equation}
which is manifestly second-order in the perturbations. The gravitational potential energy is, expanded out to second order in subscript-1 quantities (the reason for which will become clear),
\begin{equation}
    \mathcal{V} = -\frac{GM_{\rm c}^2}{R_{\rm 0, c}}\int_0^{1}\frac{m_0}{\xi_{\rm b}}\left(1-\frac{\xi_1}{\xi_{\rm b}}+\frac{\xi_1^2}{\xi_{\rm b}^2}\right)g_0 \xi_0^2d\xi_0.
\end{equation}
There is thus a first-order correction to the gravitational potential energy of the star, which varies sinusoidally in time. However, we can also compute the analogous expression for the internal energy of the star, and -- by using Equation \eqref{xibeq} for the background state -- the first-order term exactly cancels the first-order term in the gravitational potential energy. Therefore, the individual perturbations to the gravitational and thermal energy of the star are first order, but their sum is manifestly second order. We can thus calculate the change in the energy of the star, which is fundamentally second-order in the perturbations, without directly calculating the second-order perturbations to the Lagrangian position of a fluid shell (i.e., $\xi_2$ in the series $\xi = \xi_{\rm b}+\xi_1+\xi_2$). 

\subsection{Results}
\label{sec:applications}

\begin{figure*}
    \includegraphics[width=0.49\textwidth]{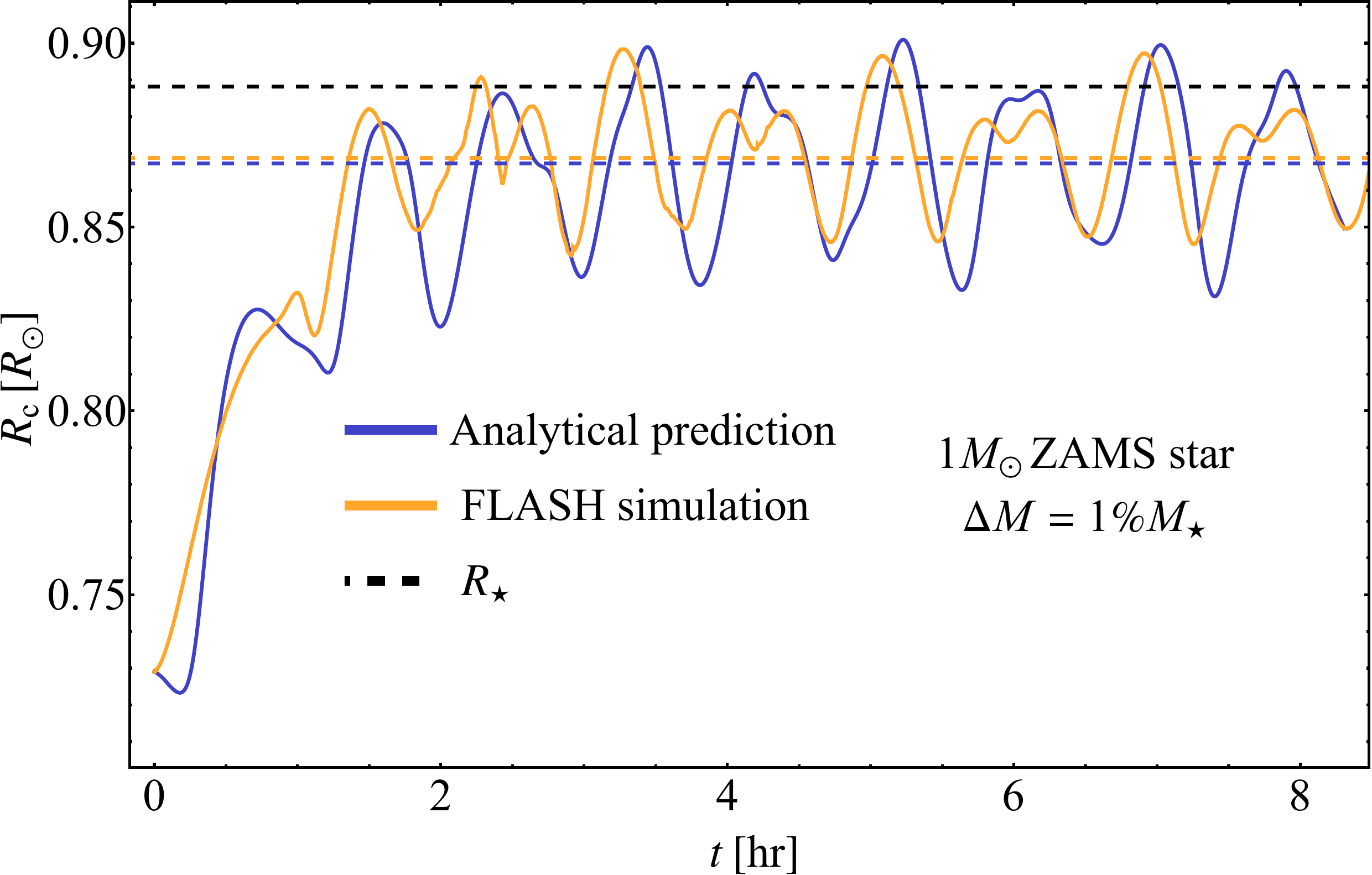}
    \includegraphics[width=0.48\textwidth]{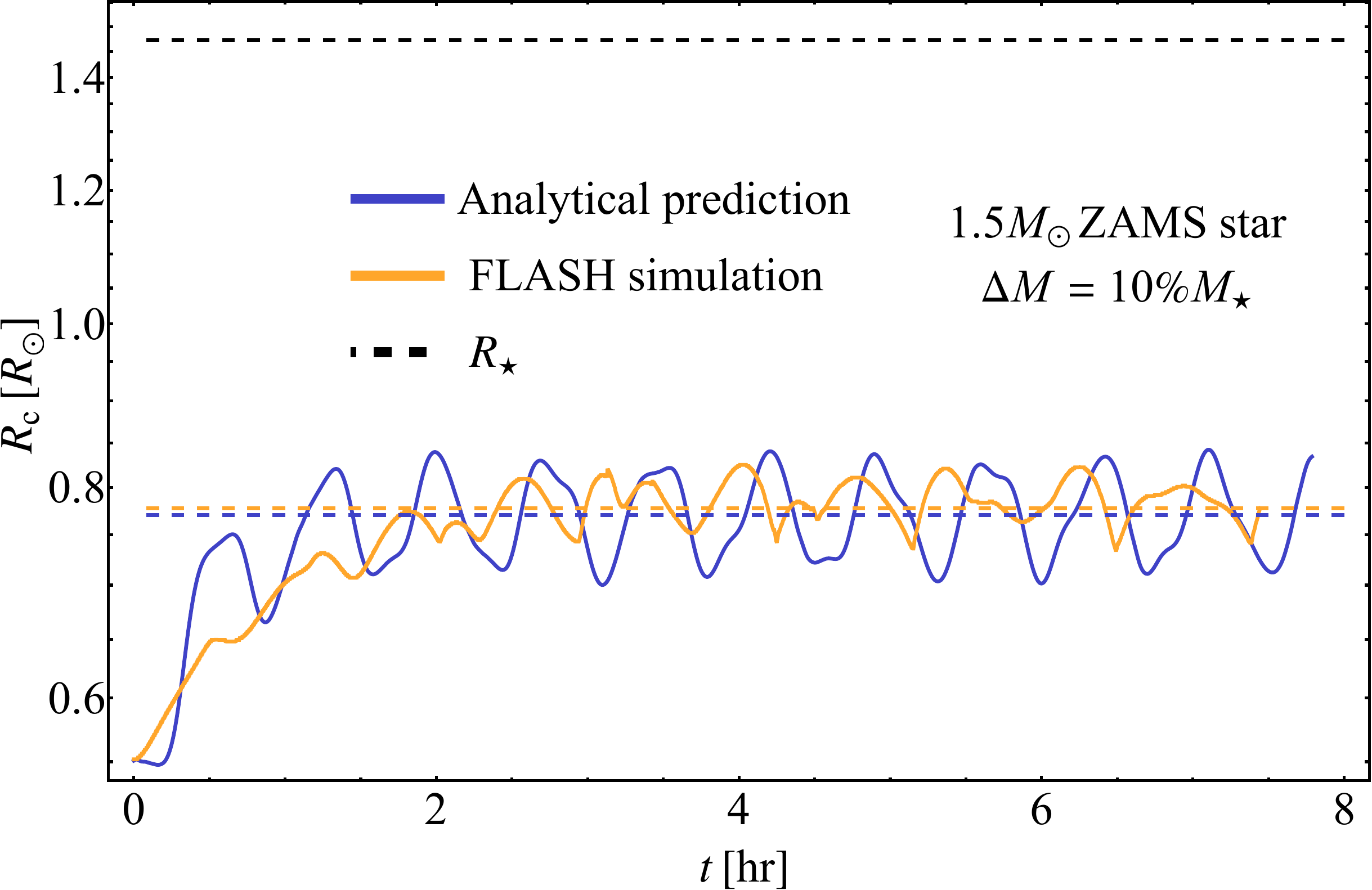}
    \caption{The time-dependent radius of the core of a $1 M_{\odot}$ ZAMS star, following a $1\%$ mass-loss (left) and a $1.5 M_{\odot}$ ZAMS star, following a $10\%$ mass-loss (right). In each case, the solid blue curve shows the analytical solution, and the solid orange curve shows the result of a hydrodynamical simulation performed using {\sc flash}. The dashed blue and orange curves indicate the average radius about which the surface oscillates, for the analytical solution and the {\sc flash} simulation, respectively. The black dashed line indicates the original radius of the star. } \label{fig:ZAMS1_radius}
\end{figure*}

Here we apply the preceding formalism to {\sc mesa}-generated stellar models. The left panel of Figure~\ref{fig:ZAMS1_radius} shows the time-dependent response of the the core radius (i.e., the total solution, $R_\mathrm{c}  = R_{0,\mathrm{c}}(\xi_\mathrm{c}+\xi_1)$) for a $1M_{\odot}$ ZAMS star following the removal of $1\%$ of its mass, where the blue curve is the model prediction. To assess the accuracy of the model, we also simulated the mass-loss process using the finite-volume hydrodynamics code {\sc flash} (V4.7; \citealt{fryxell00}). The core of the star was mapped onto a uniform spherical grid comprised of $2^{14}=16384$ cells~\footnote{We also performed the {\sc flash} simulations at a resolution of 8192 cells, to verify that our results are not sensitive to the numerical resolution.}. We used an adiabatic equation of state with $\Gamma=5/3$, and the self-gravity of the core was included. We imposed a reflecting boundary condition for all the fluid variables at the inner edge of the domain, as well as for the velocity and density at the outer boundary. To simulate the mass-loss and subsequent pressure de-confinement of the core, the pressure at the outer edge of the domain was reduced from its initial value at $R_{\rm 0,c}$ to $\sim$ zero as $e^{-\tau}$, i.e., $\omega = 1$. The orange curve is the result obtained from {\sc flash}, and is in very good agreement with the analytical solution. The average radius of the core (depicted by the dashed blue/orange line for the analytical/numerical model) is $\sim 0.867 R_{\odot}$, which is slightly smaller than the original stellar radius $R_{\star} = 0.888 R_{\odot}$, indicating that the core has a higher average density compared to the original star (despite the reduction in its mass). 

The right panel of Figure~\ref{fig:ZAMS1_radius} shows a comparison of the time-dependent radial oscillations predicted by the analytical model and a {\sc flash} simulation for a $1.5M_\odot$ ZAMS star from which $10\%$ by mass of its outer envelope was removed. For this star, $R_{\rm 0,c}\approx 0.55 R_{\odot}$ for a $10\%$ mass-loss. The mean radius about which the star oscillates is $\langle R_{\rm c}\rangle_{\mathrm{t}} \approx 0.77 R_\odot$ as predicted by the analytical model, and $\langle R_{\rm c}\rangle_\mathrm{t} \approx 0.78 R_\odot$ from the {\sc flash} simulation. Since the original radius of the star is $R_\star \approx 1.5 R_\odot$, the time-averaged radius of the core following mass loss is significantly smaller than $R_\star$, suggesting that the star undergoes a substantial increase in its average density following mass loss.

\begin{figure}
    \includegraphics[width=0.49\textwidth]{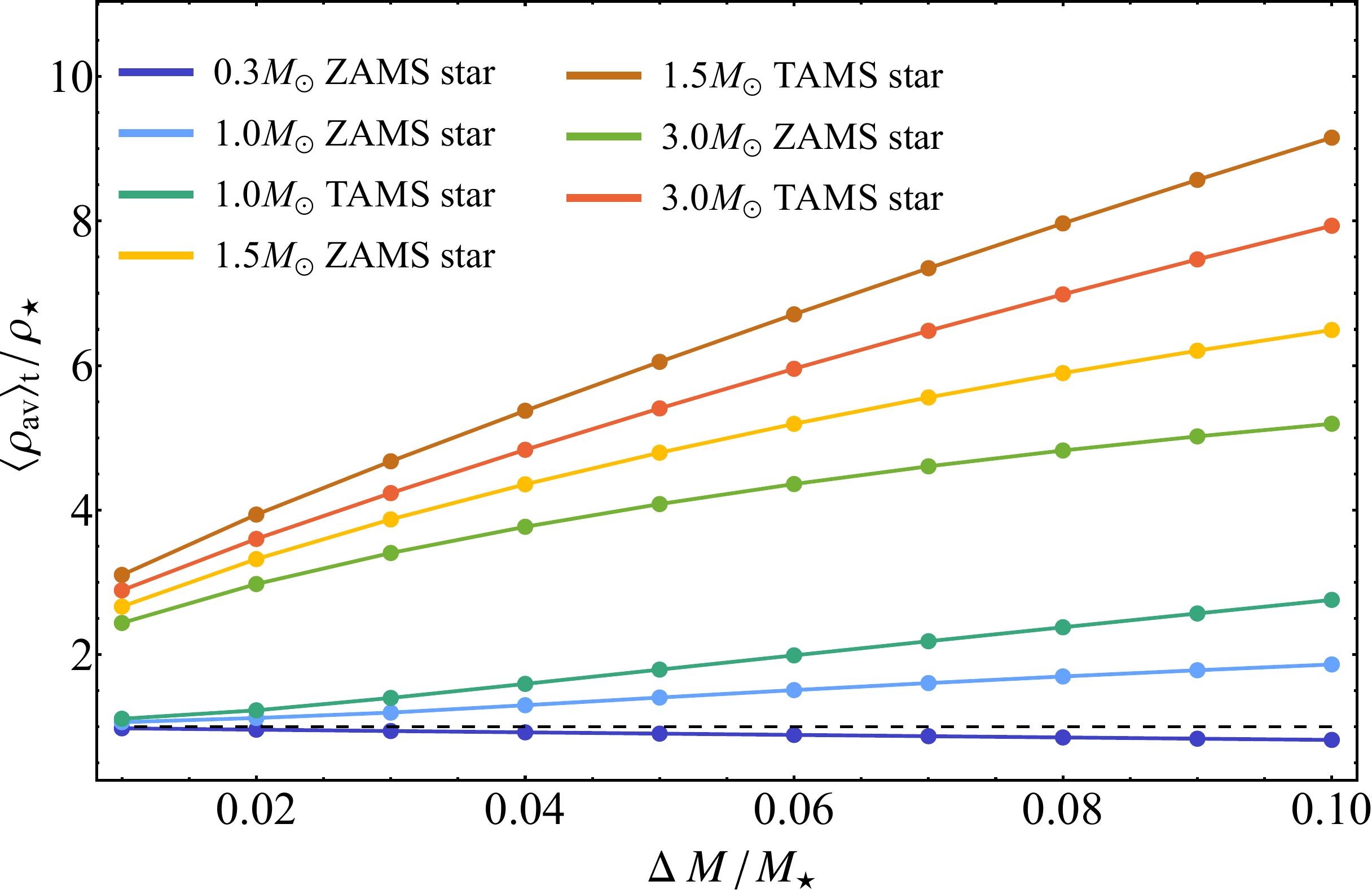}
    \caption{The time-averaged density of a star post-mass-loss, normalized by the average density of the original star, for the range of stars in the legend. The black dashed line demarcates $\langle \rho_{\rm av} \rangle_\mathrm{t}/\rho_\star = 1$.} \label{fig:mesa_avg_densities}
\end{figure}
\begin{figure*}
    \includegraphics[width=0.49\textwidth]{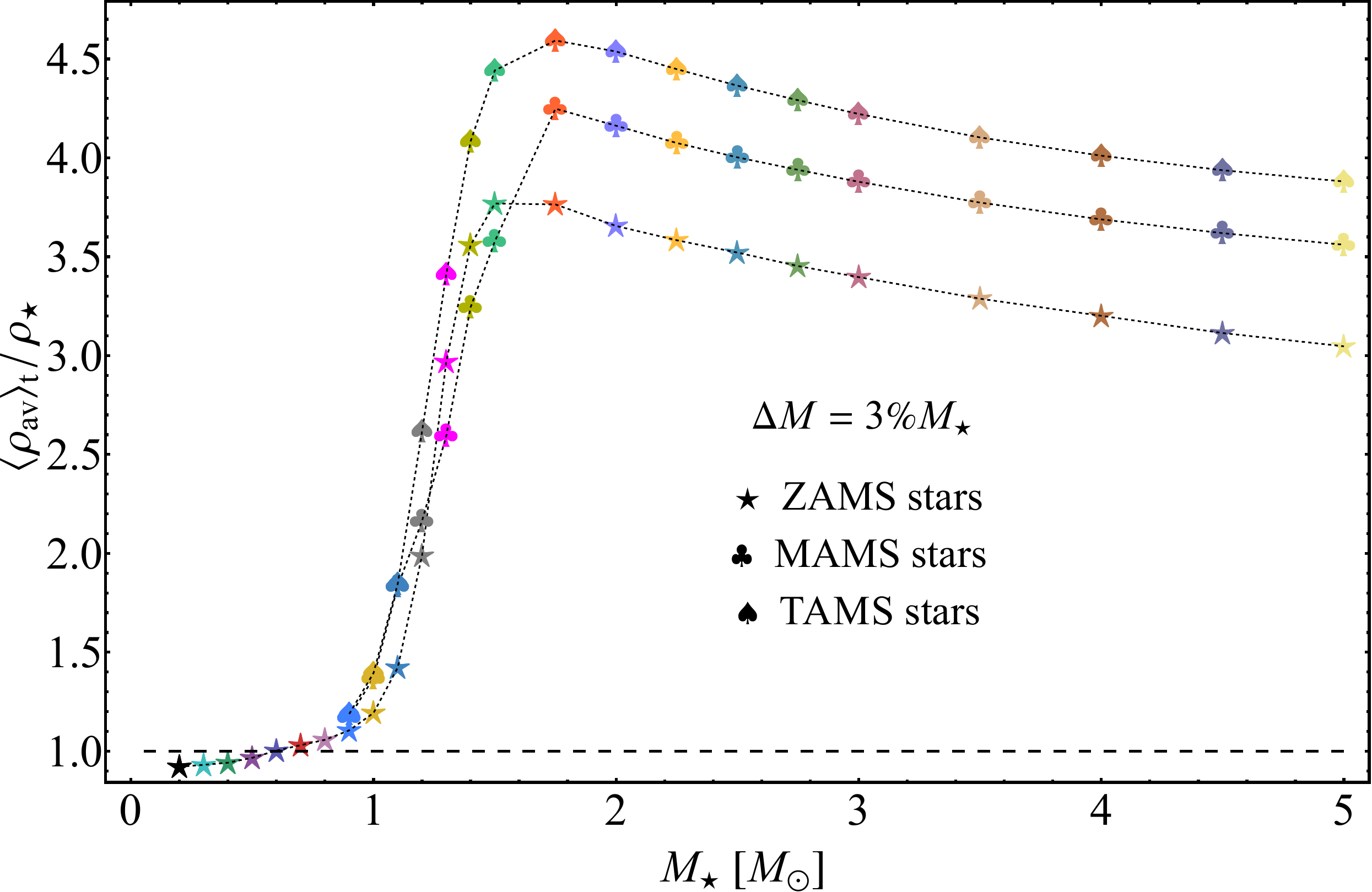}
    \includegraphics[width=0.48\textwidth]{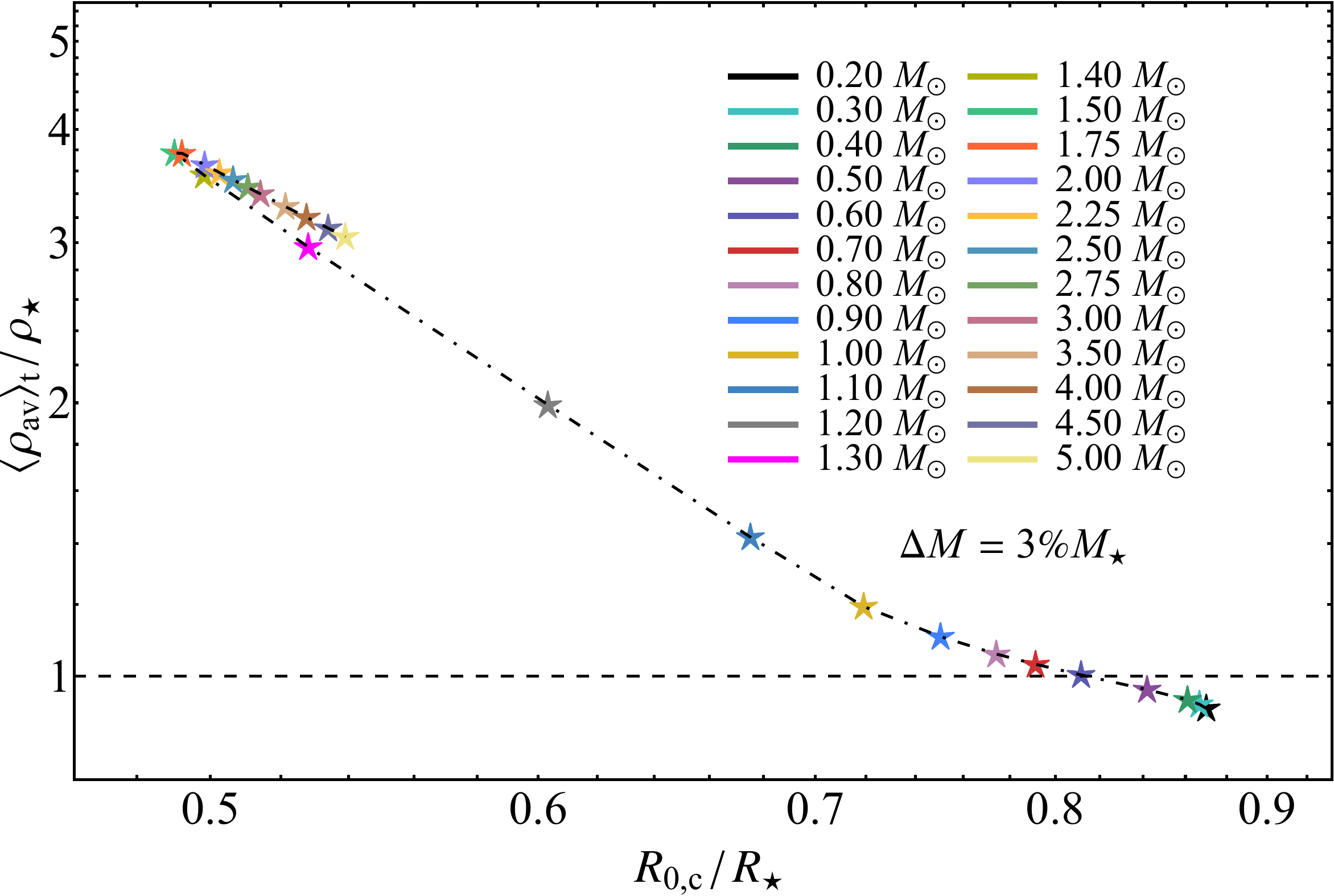}
    \caption{Left: the relative change in the average density with respect to the original average density, $\langle \rho_{\rm av}\rangle_{\rm t}/\rho_\star$, as a function of stellar mass, $M_\star$, for a $\Delta M = 3\%$ mass loss. The stars, clovers and spades represent ZAMS, MAMS and TAMS stars respectively. Right: the relative change in the average density for the ZAMS stars shown in the left panel, as a function of the initial core radius $R_\mathrm{0,c}$ (the radius in the original star that encloses $97\%$ of its mass) normalized by the stellar radius $R_\star$. The stellar masses are shown in the legend.} \label{fig:mesa_stars_density_trend}
\end{figure*}

\begin{figure}
    \includegraphics[width=0.495\textwidth]{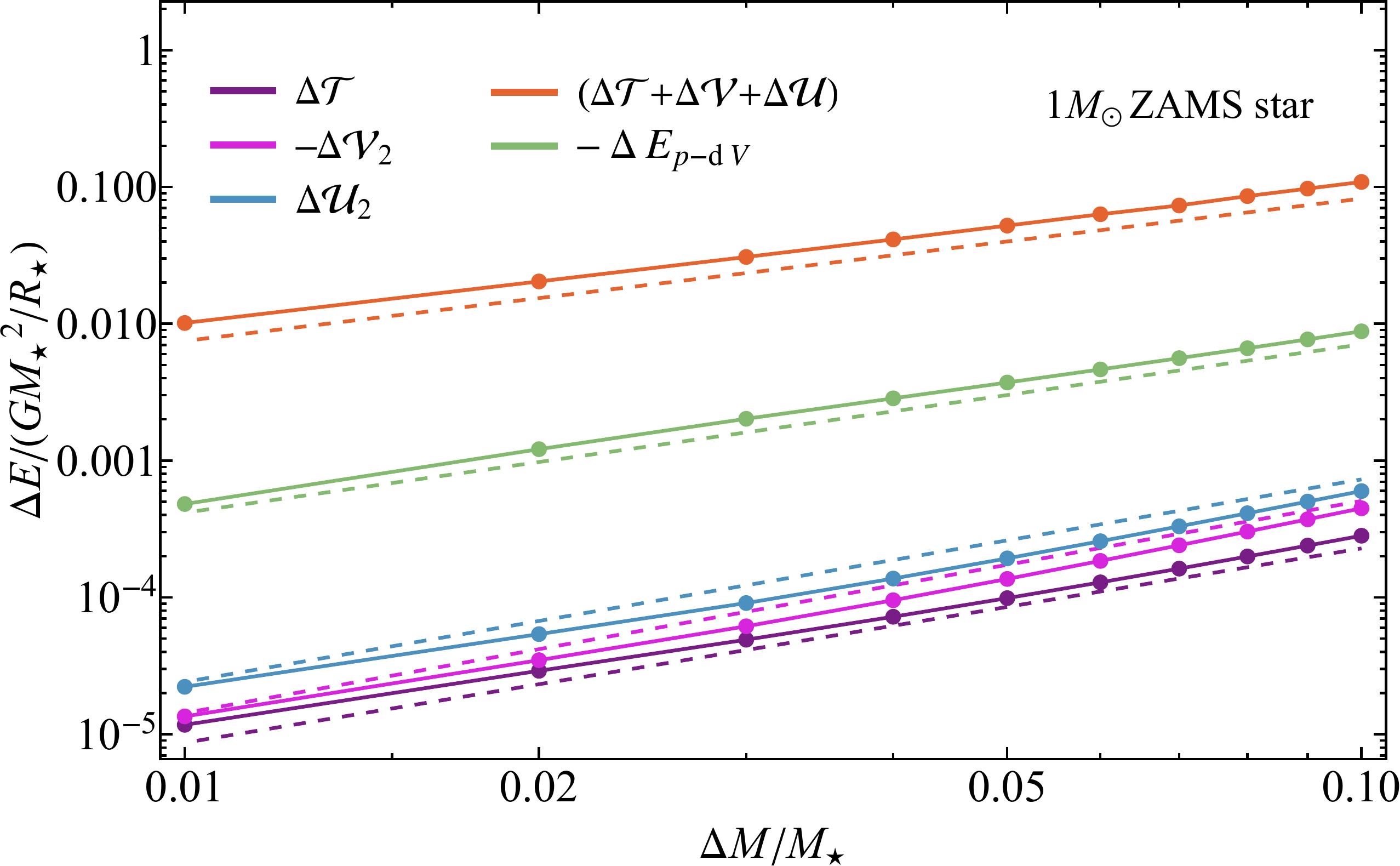}
     \caption{The time-averaged change in the kinetic energy (violet), the total energy (orange) and the $p-\mathrm{d}V$ work done by the expansion of the core (green) for a $1M_{\odot}$ ZAMS star as a function of the relative mass loss. Also shown are the second-order changes to the potential energy (magenta) and the internal energy (blue). The total change in the energy of the star is primarily determined by the difference in the binding energy of the core pre-mass-loss and that of the entire star. The dashed lines indicate approximate power-law scalings, which are also listed in Table~\ref{tab:table1}.} 
     \label{fig:average-energies}
\end{figure}

Figure~\ref{fig:mesa_avg_densities} shows the spatially and temporally averaged density $\langle \rho_{\rm av} \rangle_{\rm t}$, normalized by the average stellar density $\rho_{\star} = 3M_{\star}/(4 \pi R_{\star}^3 )$, as a function of the relative mass loss for the stellar profiles indicated in the legend. For the $0.3M_\odot$ star, which has a main-sequence lifetime longer than the age of the universe, we only include the ZAMS stage. For the higher-mass stars, we show the change in density as a function of mass lost for the ZAMS and TAMS (terminal-age main sequence) stars. As seen in the figure, the average density of the $0.3 M_\odot$ ZAMS star is lower than its original density for all values of $\Delta M / M_\star$, and decreases monotonically with mass lost. The average density of the $1M_\odot$ star, as well as that of the higher mass ones is a monotonically increasing function of $\Delta M/M_\star$ for a $1-10 \%$ mass loss, and for a given stellar mass, a more pronounced change is observed for the highly evolved TAMS stars. 

For a given value of $\Delta M/M_\star$, the change in the average density of the surviving core is not a monotonic function of stellar mass, but rather depends on the location of $R_{\rm 0, c}$ relative to the original radius of the star, $R_\star$. The left panel of Figure~\ref{fig:mesa_stars_density_trend} shows the time-averaged change in the density of the core normalized by the original density of the star, $\rho_\star$, for $\Delta M = 3\% M_\star$, as a function of the stellar mass. We use different symbols to represent three different ages along the main sequence, as follows. The ZAMS stars (having a core hydrogen fraction of $\sim 0.7$) are depicted by $5-$pointed stars in the figure. The clovers and spades represent MAMS (middle-age main sequence, determined by a core hydrogen fraction of $\sim 0.2$) and TAMS (terminal age main sequence, when the core hydrogen fraction drops below $0.001$) stars respectively. Generally, stars that are further along their main-sequence evolution develop a centrally concentrated core, such that the removal of a small amount of mass yields a smaller value of $R_{0,\mathrm{c}}/R_\star$ (as seen in the figure, the trend is reversed for the $1.3-1.5M_\odot$ stars at their ZAMS and MAMS stages), and a larger relative increase in the density. The right panel of Figure~\ref{fig:mesa_stars_density_trend} further illustrates the correlation between the relative change in the average density and the ratio $R_{0,\mathrm{c}}/R_\star$. The value of $R_{0,\mathrm{c}}/R_\star$ monotonically decreases with an increase in stellar mass from $0.2-1.5M_\odot$, where the relative change in the average density is maximized. Beyond $M_\star = 1.5 M_\odot$, removal of $3\%$ of the star's mass results in a diminishing value of $R_{0,\mathrm{c}}/R_\star$, and consequently yields a less pronounced increase in the average density of the core. For the small fractional mass losses ($\Delta M \sim 1-10\% M_\star$) considered in this work, the change in the central density is a modest (for low-mass stars) to negligible (for high-mass and evolved stars) fraction of its original value. While the central density plays an important role in determining the distance at which a star is completely destroyed~\citep{lawsmith20,ryu20,coughlin22}, its impact on the response of stars to the small fractional mass losses considered here is insignificant, as it is largely the average density that determines the partial tidal disruption radius~\citep{guillochon13,miles20,lawsmith20,ryu20,coughlin22,jancovic23}.

\begin{table*}
\begin{center}
\begin{tabular}{|c|c|c|}
\hline
Stellar mass and age & \multicolumn{2}{c|}{Power Law Fits} \\
\cline{2-3}
& $\Delta \mathcal{T}$ & $(\Delta \mathcal{T}+\Delta \mathcal{U}+\Delta \mathcal{V})$  \\
\hline
$0.3M_{\odot}$ ZAMS star & $1.8\times10^{-2}(\Delta M/M_{\star})^{1.50}$ & $0.9\times(\Delta M/M_{\star})^{0.97}$ \\
$1.0M_{\odot}$ ZAMS star & $6.1\times10^{-3}(\Delta M/M_{\star})^{1.38}$ & $1.2\times(\Delta M/M_{\star})^{1.04}$ \\
$1.0M_{\odot}$ TAMS star & $4.2\times10^{-3}(\Delta M/M_{\star})^{1.28}$ & $1.4\times(\Delta M/M_{\star})^{1.08}$ \\
$1.5M_{\odot}$ ZAMS star & $1.3\times10^{-2}(\Delta M/M_{\star})^{1.55}$ & $1.9\times(\Delta M/M_{\star})^{1.09}$ \\
$1.5M_{\odot}$ TAMS star & $8.7\times10^{-3} (\Delta M/M_{\star})^{1.41}$ & $2.1\times(\Delta M/M_{\star})^{1.11}$ \\
 \hline
\end{tabular}
\caption{\label{tab:table1} Power-law scalings for the change in kinetic energy $\Delta \mathcal{T}$ and total energy $(\Delta \mathcal{T}+\Delta \mathcal{U}+\Delta \mathcal{V})$ with fractional mass removed for the stars in the left column.}
\end{center}
\end{table*}
Figure~\ref{fig:average-energies} shows the temporally averaged change in the total energy of the star as a function of the fractional mass lost, for the $1M_\odot$ ZAMS star. Also shown are the contributions from the kinetic energy (which is manifestly second order in the perturbation $\xi_1$), as well as the potential and internal energy terms of $\mathcal{O}(\xi_1^2)$ and the $p-\mathrm{d}V$ work done by the expansion of the core. The change in the energies are plotted relative to the energy of the original star, and normalized by $G M_\star^2/R_\star$. The energies scale approximately as power-laws in $\Delta M/M_{\star}$. Table~\ref{tab:table1} lists the approximate power-law scalings, for a range of stellar masses and ages. Irrespective of stellar type, the kinetic energy of the oscillatory modes is several orders of magnitude smaller than the net change in energy, as well as the zero-order changes in the potential and internal energy terms (not shown in the figure). This shows that the conventional ``tidal heating'' term is subdominant to the change in the gravitational and thermal energies, and has a negligible role in determining the total energy balance of the surviving stellar core. The net change in energy is positive, which, as noted in Section~\ref{sec:energy}, indicates that the final state of the star is slightly less gravitationally bound than the original star, despite the core being more bound to itself as a result of the mass loss. 

\section{Comparison with numerical simulations of TDEs}
\label{sec:phantom} 
The preceding formalism yields an analytical methodology for understanding the adiabatic response of any star to mass loss, the results of which have implications for repeating partial TDEs (see Section \ref{sec:implications} below). However, this numerical simplicity comes at the expense of ignoring nonlinear terms, and we also did not account for the rotation that is imparted to the star following the partial disruption, which is on the order of tens of percent the breakup velocity \citep{bandopadhyay24}. More generally, our claim is that the primary response of the star to tidal stripping is determined by the amount of mass lost, but there are obvious differences between our simple model -- effectively a Bonnor-Ebert sphere that undergoes pressure de-confinement -- and the nonlinear tidal interaction that removes mass from the star in a partial TDE. This begs the question of how well the preceding results even qualitatively agree with hydrodynamical simulations of partial TDEs, let alone quantitatively. 

To address these uncertainties, we simulated the partial disruption of three stars -- the $0.3M_\odot$ ZAMS, $1M_{\odot}$ ZAMS and $3 M_\odot$ TAMS stars considered in \citet{golightly19} -- by a $10^6 M_{\odot}$ SMBH with the SPH code {\sc phantom} \citep{price18}. We used $10^6$ particles, and the details of the numerical method are identical to those described in~\cite{golightly19,bandopadhyay24}. For these simulations, the thermal energy generated as a byproduct of viscous dissipation is retained within the gas. The amount of mass stripped from the star is determined by the impact parameter $\beta \equiv r_{\rm t}/r_{\rm p}$, which is the ratio of the tidal radius $r_{\rm t} = R_{\star} \left( M_{\bullet}/M_{\star}\right)^{1/3}$ (where $M_{\bullet}$ is the mass of the SMBH, $M_{\star}$ and $R_{\star}$ are the mass and radius of the star respectively), to the pericenter distance of the orbit, $r_{\rm p}$. 
We use a density cutoff to delineate the core from the stream particles, such that the core is defined as the subset of particles for which $\Delta\log(\rho) <0.01$, where $\Delta \log(\rho)$ is the difference between the logarithm of the density of a particle and that of the particle having the next highest density (see Figure~3 of~\cite{bandopadhyay24} for a particle plot showing the obvious separation between the core and stream). The mass of the core, $M_{\rm c}$, is the sum of the masses of all the particles that constitutes the core, and the core radius, $R_{\rm c}$, is defined as
\begin{equation}
    R_{\mathrm{c}} =  \sqrt{x_{\rm c}^2+y_{\rm c}^2+z_{\rm c}^2} \bigg{|}_{z_{\rm c}=z_{\rm m}}, \label{eq:core-radius}
\end{equation}
where $x_{\rm c} = x-x_{\rm COM}$ etc., and $z_{\rm m}$ is the absolute value of the center-of-mass (COM) subtracted $z-$coordinate maximized over all the core particles. We note that the definition of $R_{\rm c}$ as given by Equation~\ref{eq:core-radius} is not sensitive to the exact value of $\Delta\log(\rho)$ used to determine the core, since the density gradient in the direction perpendicular to the orbital plane is much stronger than that in the orbital plane. We verified the insensitivity to the criterion used for establishing the core radius by choosing a different value, $\Delta\log(\rho) < 0.05$, which yields an almost identical estimate for $R_{\rm c}$.

\begin{figure}
    \includegraphics[width=0.48\textwidth]{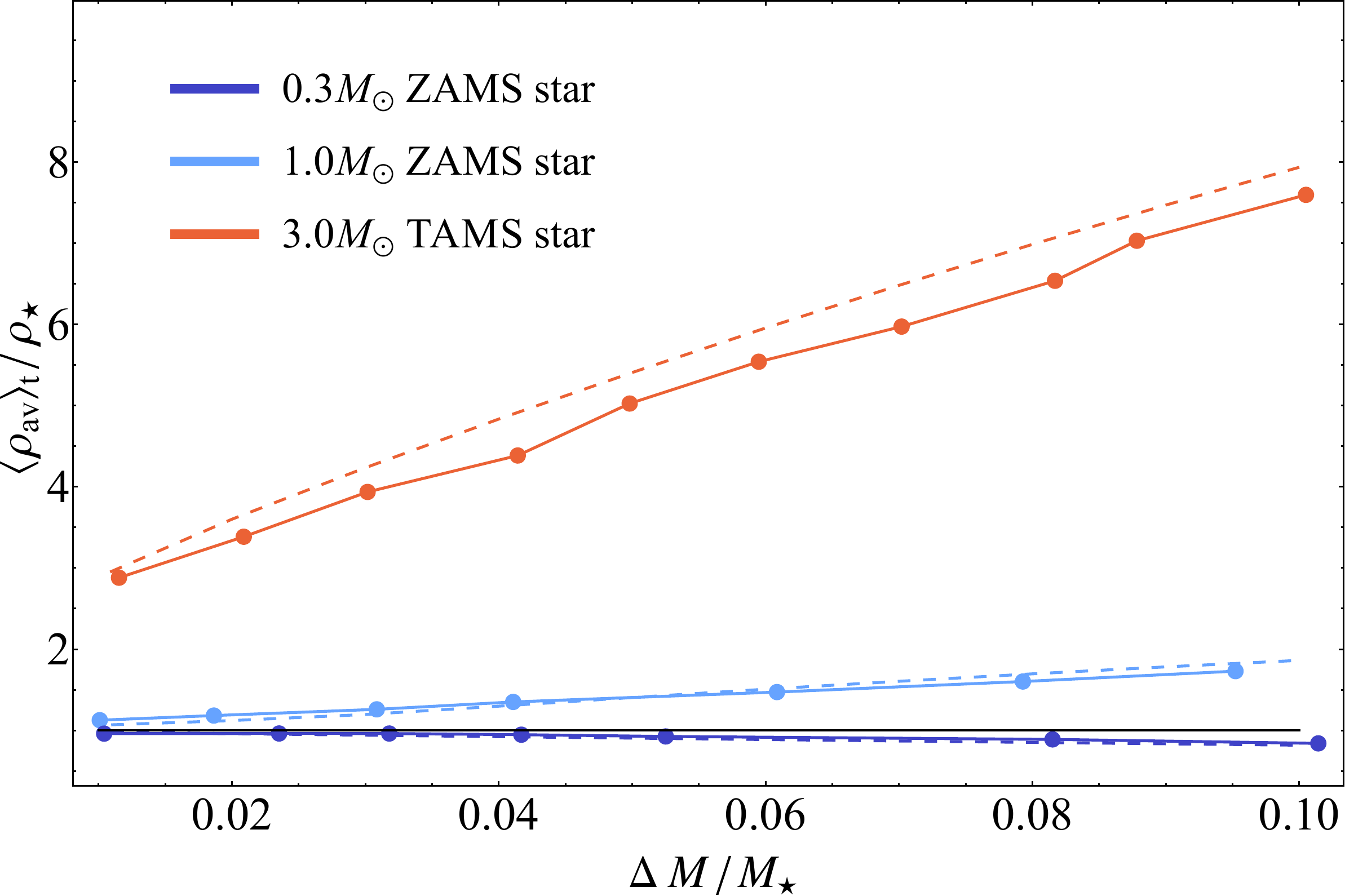}
    \caption{The average density of the surviving core, normalized by the average density of the original star, for three different stars evolved using {\sc mesa}. The solid lines depict the results of TDE simulations performed  with {\sc phantom}, where $\Delta M/M_\star$ is determined by the pericenter distance of the orbit. The dashed lines are the predictions of the analytical model. The solid black line depicts $\langle \rho_{\rm av} \rangle_\mathrm{t}/\rho_\star = 1$. For the $0.3M_\odot$ ZAMS star, the average density of the core is always less than the original density of the star, and decreases with increasing mass lost. The $1.0M_{\odot}$ ZAMS and $3.0M_\odot$ TAMS stars exhibit a monotonic increase in the average density of the core with increasing $\Delta M/M_\star$, with a more pronounced change observed for the $3M_\odot$ TAMS star.} \label{fig:phantom_densities}
\end{figure}

\begin{figure*}
    \includegraphics[width=0.495\textwidth]{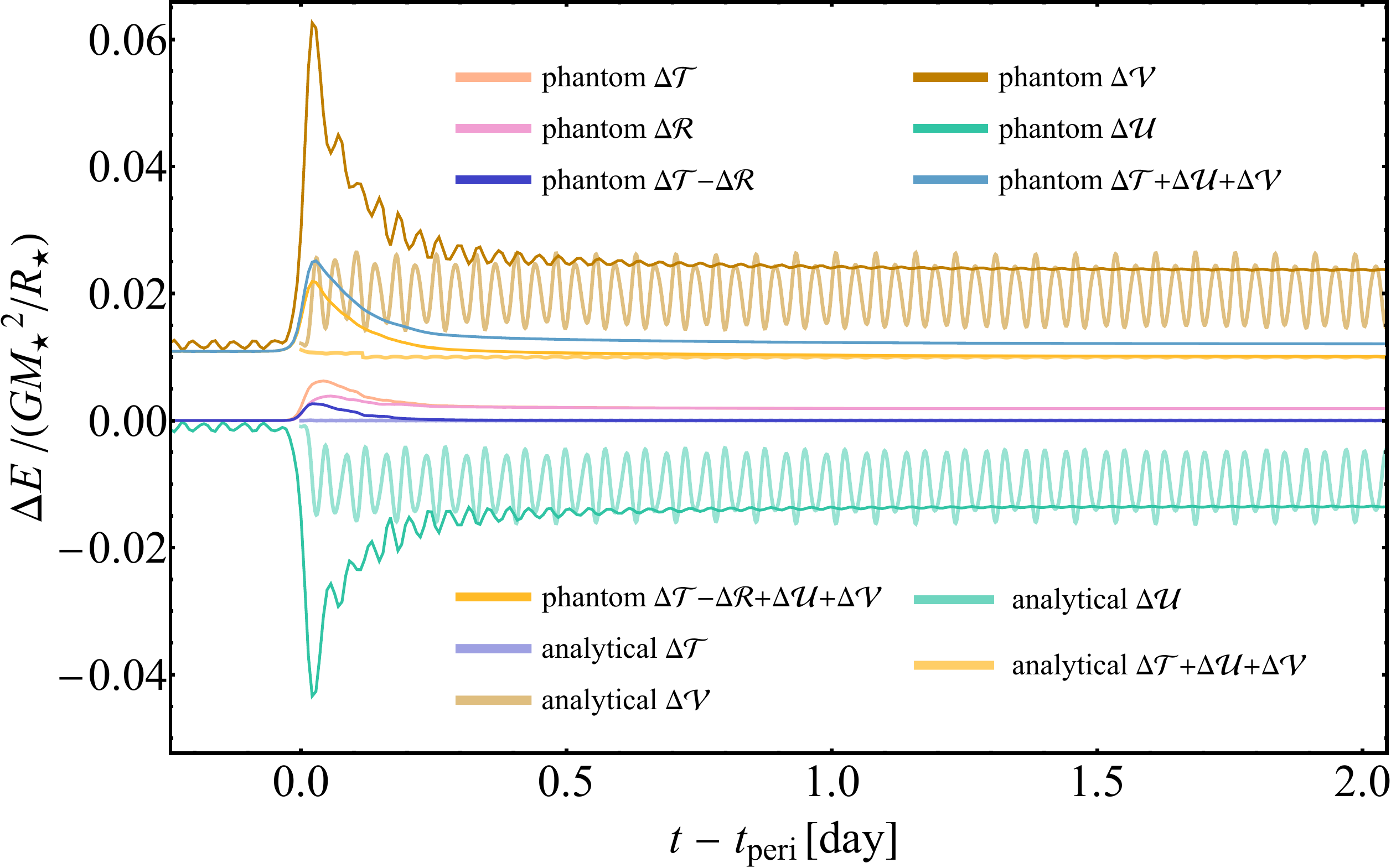}
     \includegraphics[width=0.495\textwidth]{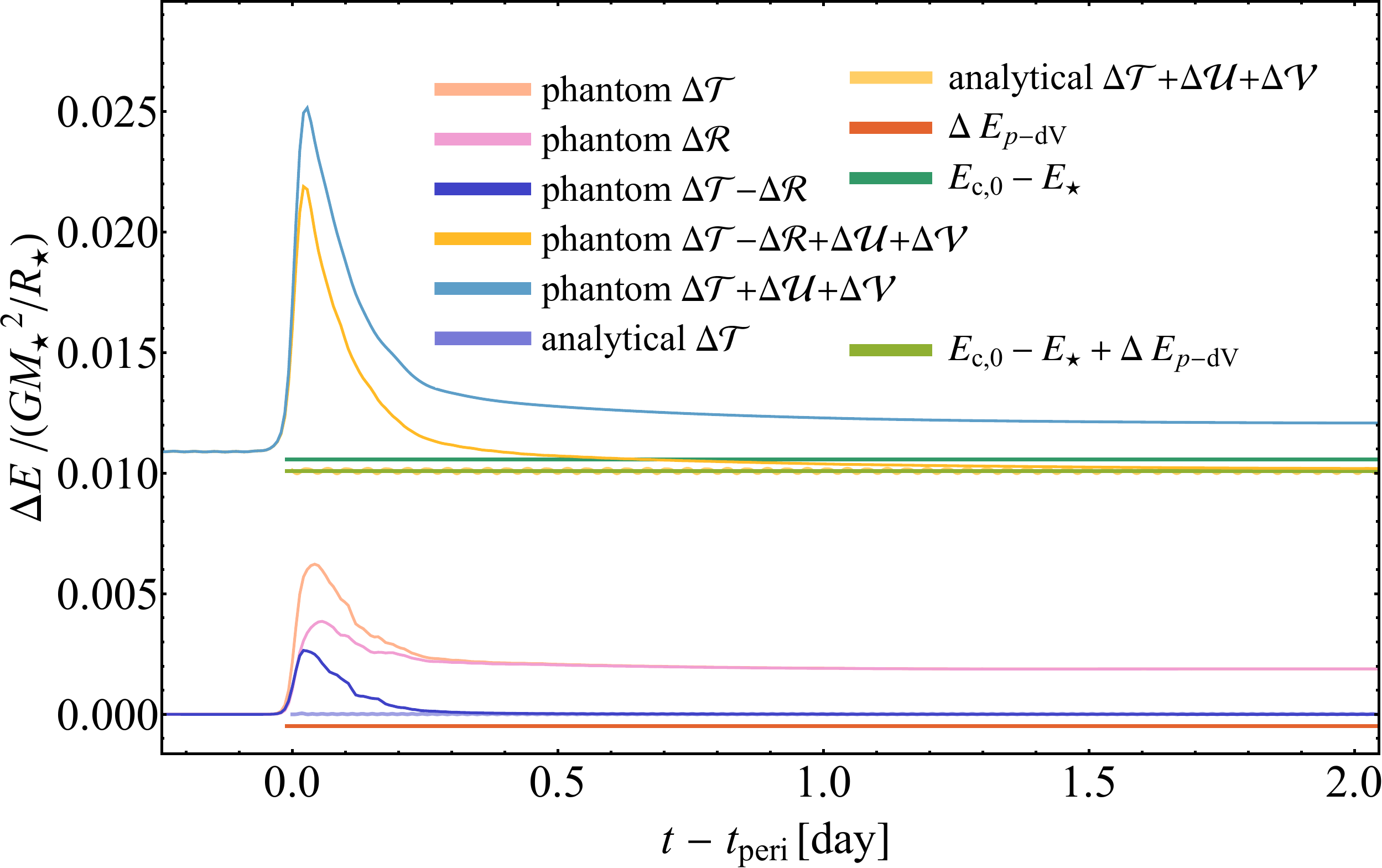} 
    \caption{Left: the change in the kinetic, rotational, potential and internal energies of the $1 M_{\odot}$ ZAMS star on a $\beta=0.7$ orbit around a $10^6 M_{\odot}$ SMBH, and the total energy imparted to the surviving core, calculated from the SPH simulation. Also shown are the predictions for the different energy contributions obtained using the analytical model. The kinetic energy and the rotational energy contributions, as well as the total change in energy, are $\sim 1-2$ orders of magnitude smaller than the individual changes in the internal and potential energies.  Right: a zoomed-in version of the left panel. We also show the change in the binding energy resulting from mass loss, $E_{\rm c,0}-E_\star$, the work done by the expansion of the core, $\Delta E_{p-\mathrm{d}V}$, and their sum, which is comparable to the total change in the core energy.}
    \label{fig:ZAMS1-sph-energies}
\end{figure*}

Figure~\ref{fig:phantom_densities} shows the corresponding density\footnote{We use the same notation as the analytical section, namely that the density is ``time-averaged,'' but we note that the core has effectively reached a steady state by the time we measure it and is no longer oscillating; see, e.g., the brown and blue curves in Figure \ref{fig:ZAMS1-sph-energies} appropriate to the {\sc phantom} simulations.} of the surviving core, i.e., $\langle \rho_{\rm av}\rangle_{\rm t} = 3M_{\rm c}/(4\pi R_{\rm c}^3)$, normalized by the average density prior to mass loss, for all three stars simulated using {\sc phantom}, for a range of mass losses between $1-10\% M_\star$. We show the analytical predictions alongside the numerical results, using dot-dashed lines to depict the former. For the $0.3M_\odot$ ZAMS star, the average density decreases monotonically with mass lost, indicating that the star becomes more susceptible to mass loss on subsequent encounters with an SMBH. Conversely, the average density increases as a function of mass lost for the $1M_\odot$ ZAMS and $3M_\odot$ TAMS star, with the average density of the surviving core being $\sim 8$ times that of the original star for the $3 M_\odot$ star losing $10\%$ of its mass. This shows, in agreement with our analytical model, that the removal of small amounts ($\sim 1-10\%$) of mass from the outer envelope of massive stars stabilizes them against further mass loss, making such stars good candidates for surviving repeated tidal encounters with an SMBH.

Figure~\ref{fig:ZAMS1-sph-energies} compares the energies imparted to the surviving stellar core (where ``$\Delta$'' represents a change in the energy relative to the original star, prior to mass loss) as a function of time from the $\beta=0.7$ disruption of the $1M_{\odot}$ ZAMS star and the analytical model. For the numerical simulation, the energies are calculated as sums over all the SPH particles comprising the core. The total kinetic energy $\Delta \mathcal{T}$ is the sum of contributions from the energy of the oscillatory modes excited through tides and the rotational kinetic energy $\Delta \mathcal{R}$ (the rotational velocity of the fluid is defined relative to the COM and identically to \citealt{bandopadhyay24}), and hence the difference $\Delta \mathcal{T}-\Delta \mathcal{R}$ gives the energy contained in the oscillatory modes only, which is shown in purple in the right panel. The kinetic energy of the tidally excited modes thus rises and peaks as the star reaches pericenter, after which it decays sharply, with most of the contribution to $\Delta \mathcal{T}$ being the rotational kinetic energy at times $\gtrsim 0.5$ days post-pericenter. The coupling of the first-order tidal oscillations (in particular, the $\ell=2 , \, m=-2$ mode; see discussion in ~\citealt{kochanek92}) gives rise to bulk rotation of the stellar core, as indicated by a growth in the rotational kinetic energy. The residual $\Delta \mathcal{T}-\Delta \mathcal{R}$ (purple curve) is in good agreement with the kinetic energy of the modes calculated from the analytical model. The changes in the gravitational potential energy and the internal energy are more than two orders of magnitude larger than the kinetic energy of the oscillatory modes. The change in the gravitational potential energy always outweighs the change in the internal energy, and the sum of the two is positive at all times. As discussed in Section~\ref{sec:energy}, this implies that the core is less self-bound than the original star, as a result of mass loss. The average values of the potential and internal energy terms for the analytical model shown in Figure~\ref{fig:ZAMS1-sph-energies} agree well with the numerical result.

The analytical prediction that the sum of the gravitational, thermal, and oscillatory energies is positive is consistent with that calculated from the numerical simulation, which is shown by the dark-yellow curve. However, the contribution from the rotational kinetic energy, which our model does not incorporate, shifts the total energy (the light blue curve in the right panel of Figure~\ref{fig:ZAMS1-sph-energies}) to a slightly higher value, approximately $\sim 0.015 GM_{\star}^2/R_{\star}$. This therefore affirms that the largest contribution to the change in the energy is due to the fact that the pre-mass-loss core is less gravitationally bound than the star as a whole, but the rotational energy -- which is not included in the model -- is a factor of a few more important than the $p-dV$ work done by the core as it expands.
\begin{figure*}
    \includegraphics[width=0.495\textwidth]{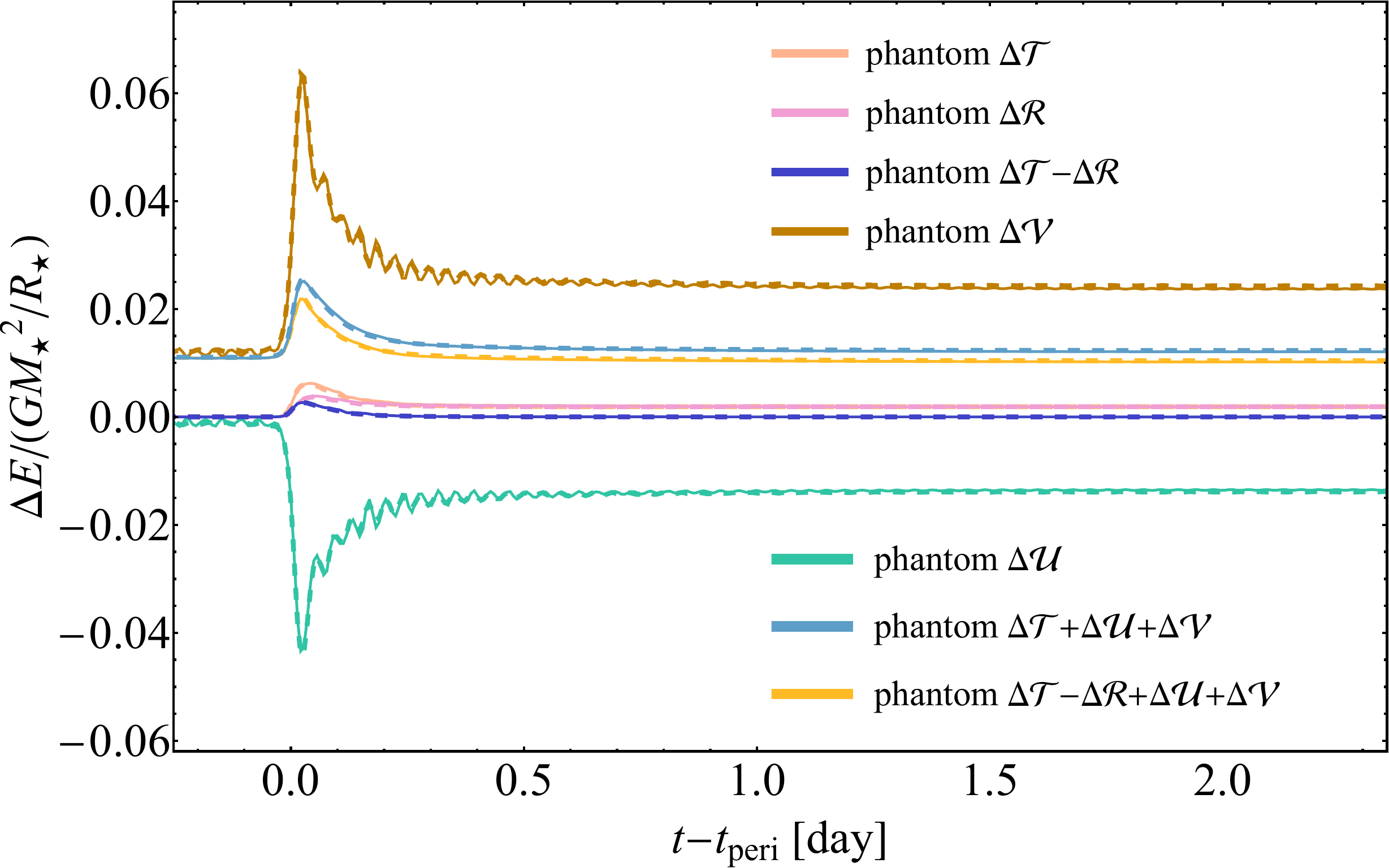}
     \includegraphics[width=0.495\textwidth]{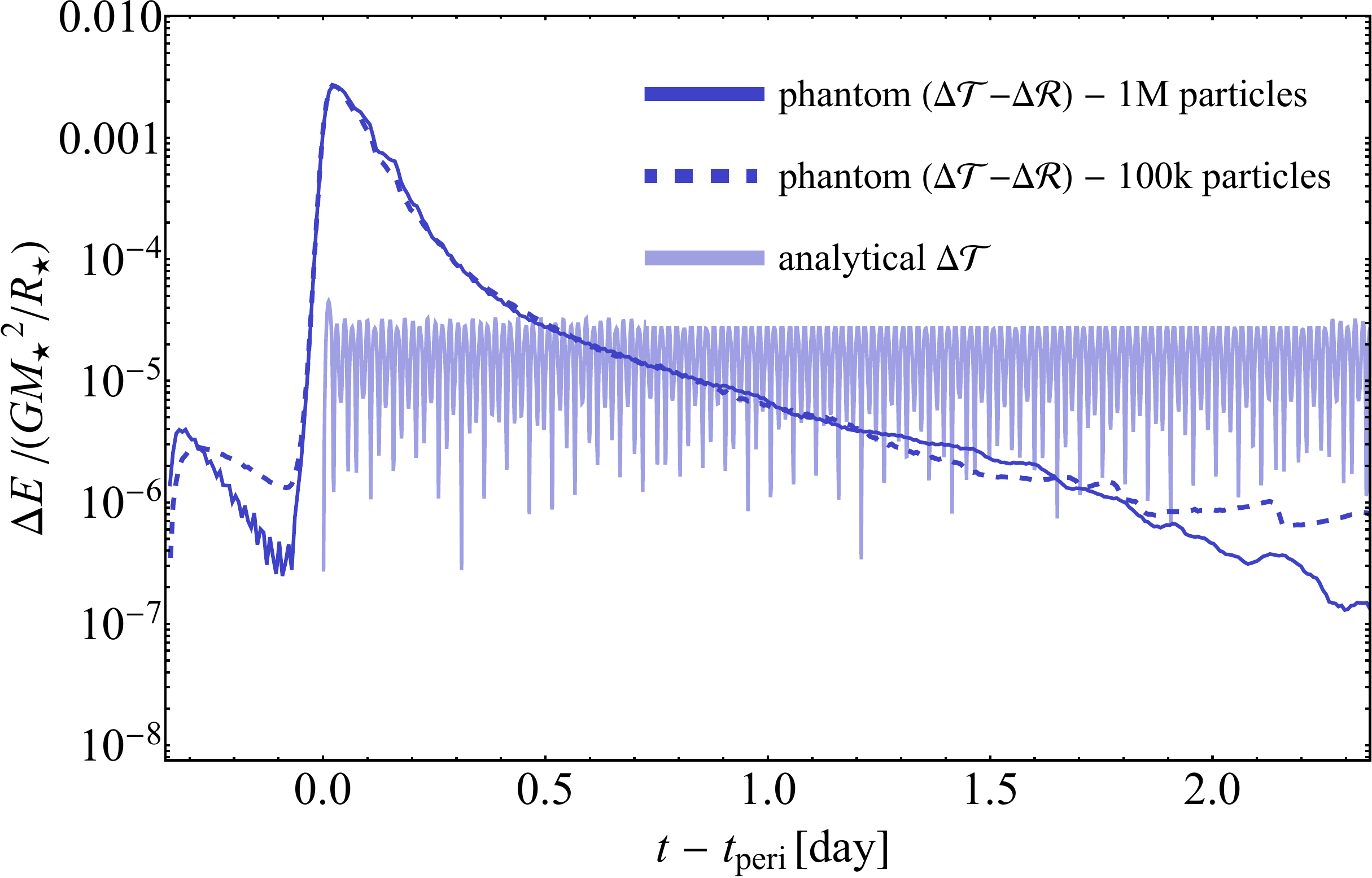} 
    \caption{Left: same as Figure \ref{fig:ZAMS1-sph-energies} but for two different SPH resolutions, where solid (dashed) corresponds to $10^6$ ($10^5$) particles; the agreement between the two illustrates the convergence of the results. Right: the kinetic energy of the oscillatory modes at the two different resolutions, as well as the analytical prediction for the kinetic energy. } \label{fig:ZAMS1-energies-convergence}
\end{figure*}

To assess the numerical convergence of the results, we also simulated the disruption of the $1M_{\odot}$ ZAMS star on a $\beta=0.7$ orbit at a resolution of $10^5$ particles. The left panel of Figure~\ref{fig:ZAMS1-energies-convergence} compares the energies at both resolutions, $10^5$ ($10^6$) particles shown by dashed (solid) curves, and they are effectively indistinguishable. The right panel shows the kinetic energy of the oscillatory modes (i.e., $\Delta \mathcal{T}-\Delta \mathcal{R}$ from the {\sc phantom} simulations) from the simulations and as predicted analytically. The fact that the two resolutions agree with one another post-pericenter\footnote{The disagreement pre-pericenter is likely numerical, and due to the fact that the changes in the kinetic energy are so small prior to pericenter that they are not numerically resolved.} strongly suggests that the decay in the oscillatory kinetic energy is not numerical, but is the result of the coupling described in \citet{kochanek92}. Beyond two days the oscillatory kinetic energy has decayed to the point that it is no longer well resolved. 

While we do not show them here, the energies for other stars and mass loss are similar: the predictions of the analytical model as concerns the gravitational, thermal, and oscillatory kinetic energies are upheld, such that the sum of all three is a small positive fraction ($\lesssim 1-10\% \,  G M_\star^2/R_\star$) of its binding energy, and the rotational energy imparted to the surviving core shifts the total energy to a slightly higher value.

\section{Discussion and Implications for Partial TDEs}
\label{sec:implications}
\subsection{Effective tidal radius and stability of the mass transfer process}
\cite{coughlin22} argued that the ``partial disruption radius'' -- the impact parameter $\beta$ at which one expects any mass to be stripped -- is largely independent of stellar structure and given by $\beta_{\rm partial} \approx 0.6$, which agrees with numerical investigations of TDEs performed in, e.g.,~\cite{guillochon13,miles20,lawsmith20,ryu20,jancovic23}. In Sections~\ref{sec:applications} and \ref{sec:phantom}, we showed that for a low-mass star, the removal of mass from its outer layers results in a decrease in its average density, rendering such stars susceptible to runaway mass loss during repeated tidal encounters with an SMBH. Contrarily, for stars with masses $\gtrsim 0.7 M_{\odot}$, a small fractional mass loss results in a new average radius that is smaller than the original radius $R_\star$, and consequently an increased average density (despite the reduction in the mass). We showed that this effect is most significant for highly evolved stars, which develop a centrally concentrated core and a diffuse outer envelope, such that for the $1.5 M_\odot$ and $3 M_\odot$ TAMS stars, a removal of $1-10\%$ of mass from their outer layers increases the average density by a factor of $\sim 3-8$.~\cite{yao25} find a similar trend for the stability of stars undergoing mass transfer in a tidal encounter with an SMBH.

The increase in the average density of a star leads to a decrease in its tidal radius $r_{\rm t} = \left( M_{\bullet}/((4/3)\pi \rho_{\star})\right)^{1/3}$. For rpTDEs of stars on bound orbits, the pericenter distance $r_{\rm p}$ remains effectively unchanged on the second and subsequent encounters, owing to angular momentum conservation and the fact that the maximum angular momentum of the star is still a small fraction of that associated with the orbit itself (see Equation 2 of \citealt{bandopadhyay24} for the maximum change in pericenter radius). Thus, an increase in the average density of the star leads to a decrease in the effective impact parameter $\beta = r_{\rm t}/r_{\rm p}$, suggesting that the amount of mass stripped should be a decreasing function of the number of pericenter passages. This is indeed the case for high-mass and evolved stars on low-$\beta$ orbits around an SMBH, e.g., Figure 7 of \citealt{bandopadhyay24} shows that the amount of mass stripped decreases with the number of pericenter passages for a $3M_{\odot}$ TAMS star on a $\beta=1$ orbit around a $10^6 M_{\odot}$ SMBH. 

For the $1M_\odot$ ZAMS star, our model predicts a modest increase in the average density, between $\sim1-1.85$ times its original value, for the same range of mass losses. However, the tidal interaction between a star and an SMBH also imparts prograde angular momentum to the star, which has a counter-balancing effect on the tidal radius~\citep{golightly19a}, and this effect is not taken into account in our analytical model. If the corresponding effect of the tidal spin-up is more important than the modest increase in density, then we would expect the star to lose more mass per encounter. This expectation is consistent with results in \citet{bandopadhyay24}, who showed that a $1M_{\odot}$ ZAMS star loses more mass on each encounter, with the fallback rate exhibiting an increasing peak magnitude (per encounter; see their Figure 13).

Finally, low-mass stars ($\lesssim 0.7 M_{\odot}$), such as the $0.3 M_\odot$ ZAMS star hydrodynamically considered here, undergo a decrease in their average density in response to mass loss. The decline in the average density -- in conjunction with the tidal spin-up -- renders such stars highly susceptible to complete disruption over a few orbits in an rpTDE. We would therefore not expect such systems to be able to power rpTDEs with many recurrent flares (e.g., ASASSN-14ko; \citealt{payne21}), or QPEs in the context of the orbiting white dwarf model (effectively $5/3$-polytropes and thus with structures analogous to low-mass stars) proposed by \citet{king20}, unless some mechanism were able to act in tandem to increase the pericenter distance per encounter. On the other hand, low-mass stars undergoing small fractional mass losses are ideally suited to reproduce the lightcurve of events such as AT2020vdq~\citep{somalwar23}, for which two flares have been observed with the second brighter than the first, i.e., a larger amount of mass could have been stripped on the second encounter.

\subsection{Energetics of the surviving core}
In Section~\ref{sec:phantom} we showed that for a mass loss of $\sim 1-10\% M_{\star}$, the kinetic energy of the oscillatory modes (which is subsequently dissipated as heat, either radiatively or through nonlinear couplings) is $\gtrsim 2-3$ orders-of-magnitude smaller than the absolute value of the change in the gravitational potential energy or the internal energy of the star (see Figure \ref{fig:ZAMS1-sph-energies}). The total change in energy (calculated as the sum of the kinetic, potential and internal energies, or using the p-dV work), is positive, and $\sim 1-10\%$ of the binding energy of the star (as shown in Figure~\ref{fig:average-energies}). These predictions agree with numerical simulations -- Figure~\ref{fig:ZAMS1-sph-energies} corroborates the model predictions, showing that the kinetic energy in the oscillations is orders of magnitude smaller than the change in the potential energy and the absolute value of the change in the internal energy. 

In the standard tidal dissipation picture, non-radial oscillatory modes are excited in a star at the expense of its orbital energy. This leads to tidal heating, which causes the star to puff up, rendering it more susceptible to further mass stripping and subsequent and unstable mass transfer~(e.g., \citealt{ray87,mcmillan87,gu04,linial24,koenigsberger24}). Our model and hydrodynamical simulations show that this picture apparently does not extend to the limit where the star 
loses a small fraction of its mass from the tidal interaction. Specifically, our results show that the thermal energy of the star decreases due to adiabatic expansion, and the kinetic energy associated with the oscillations is orders of magnitude smaller than the total (i.e., the sum of the changes to the thermal and gravitational potential energies) change in energy of the surviving core. The largest individual contributions to the change in the energy of the star arise from the gravitational and thermal energy, where the former (latter) increases (decreases) following the expansion of the de-pressure-confined surviving core, but the sum of the two (which is the net change and equal to the $p-dV$ work) is considerably smaller than the difference in the binding energy between the pre-mass-loss core and the entire star.

The tidal interaction with the black hole also spins up the stellar core, which our model neglects. Our hydrodynamical simulations show that the rotational energy of the core has a positive contribution to the total energy, but is small relative to the difference between the energy of the core pre-mass-loss and the total stellar binding energy. The total change in energy of the surviving core is therefore a small positive fraction ($\lesssim 10\%$) of the binding energy of the star, and this increase is not an effect of tidal heating, but rather is due (primarily) to the difference in the binding energy of the core pre-mass-loss and that of the original star. 
\begin{figure}
    \includegraphics[width=0.48\textwidth]{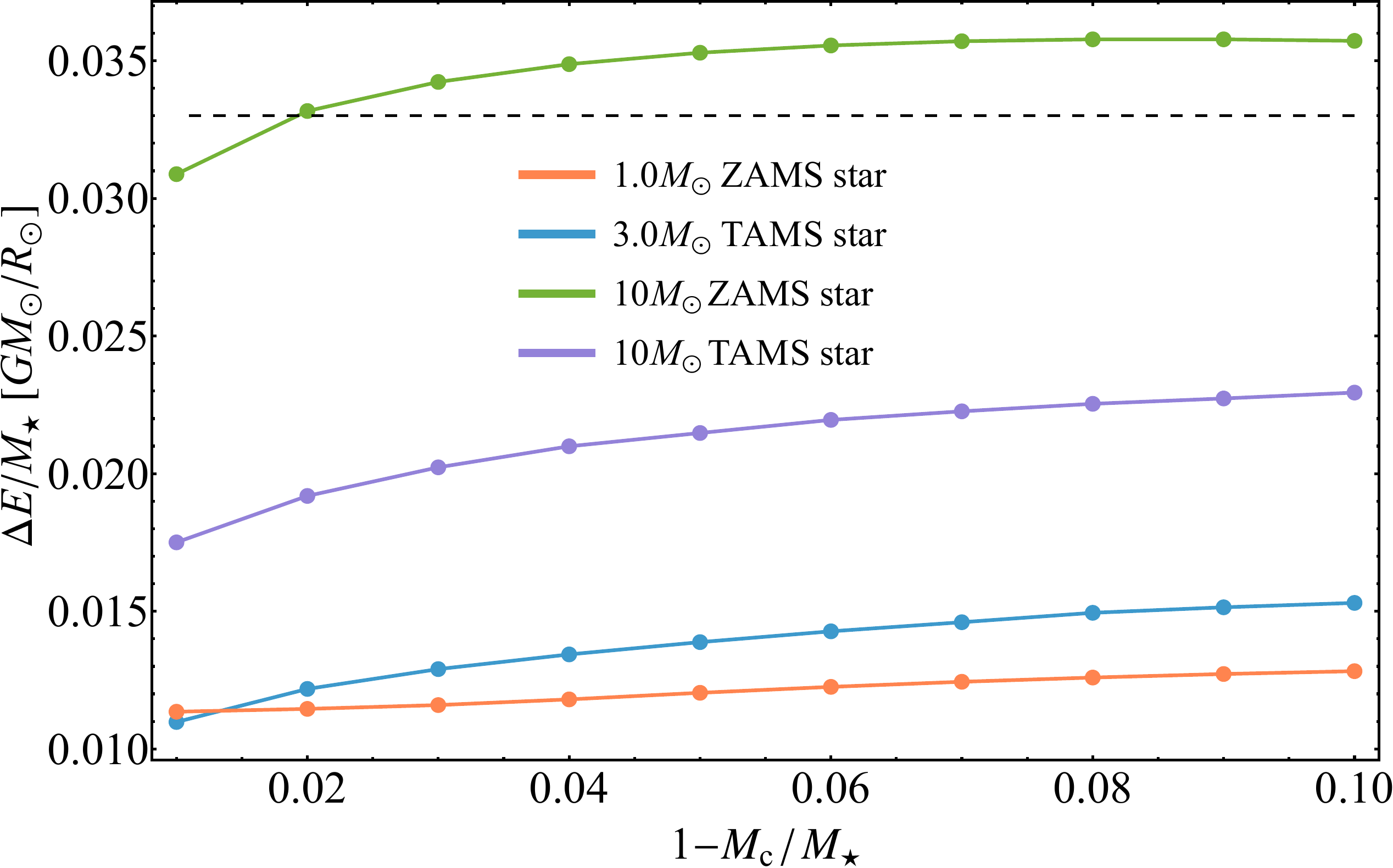}
    \caption{The per-orbit change in the specific energy of the core, in units of $G M_\odot / R_\odot$, as a function of the total fractional mass lost, for stars losing $1\%$ of their original mass per encounter with an SMBH. The black dashed line depicts the required per-orbit change in specific energy for a period decay rate of $-0.001$, assuming a $114$-day orbit about a $10^6M_\odot$ SMBH.} \label{fig:deltae-per-orbit}
\end{figure}

These results also suggest that the binding energy of the envelope could act as an additional sink of orbital energy: while the mass loss is put in by hand in our analytical model (and 1D simulations), the tidal field must do work on the star to remove this mass in the case of a TDE. If we assume that this difference in energy comes at the expense of the orbital binding energy of the star to the black hole, we can estimate the resultant change in the orbital period. Figure~\ref{fig:deltae-per-orbit} shows the change in the specific energy $\Delta \epsilon = \Delta E/M_\star$ of the core between two consecutive orbits, while being repeatedly stripped of an amount $\Delta M = 1\% M_\star$ per orbit. For a $10^6 M_\odot$ SMBH and an orbital period of $114$ days, the per-orbit change in the specific energy required to yield a period derivative $\dot{P}\sim-0.001$ is $\Delta \epsilon \approx 0.03 G M_{\odot}/R_\odot$. As seen in the figure, a comparable change in $\Delta \epsilon $ can be achieved by having a high mass ($\sim 10 M_\odot$) star on an orbit around a $10^6M_\odot$ SMBH, losing $1\%$ of its mass per orbit. For a $10^7 M_{\odot}$ SMBH, the required change in the specific energy per orbit to maintain a period derivative $\dot{P} \approx -0.001$ is $\sim 0.15 G M_\odot/R_\odot$. This could possibly be achieved by a high mass star losing $\gtrsim 5\%$ of its mass per encounter.

However, the situation is likely not this simple, as some of this energy can be retained in the tidal tails; in the case of a complete disruption, for example, the binding energy of the star would imply that slightly more than half of the star would be bound to the black hole ($\sim 51\%$ for typical numbers; \citealt{rees88}). Additionally, the notion of the tidal field doing work on the star effectively treats the latter as an in-tact object that possesses internal degrees of freedom, but this schema is at odds with the fact that large mass losses result in the reformation of the star well past pericenter \citep{guillochon13, nixon21}. If one instead assumes that the gas expands quasi-ballistically after pericenter for some amount of time before the core reforms near the location of maximum density, the core is systematically placed onto a {positive-energy} orbit owing to the difference in shear of the expanding debris \citep{coughlin25}, which agrees with hydrodynamical simulations when the mass lost is $\gtrsim 10\%$ of the mass of the original star (e.g., \citealt{manukian13, gafton15, kremer22, cufari23, kirouglu23, vynatheya24,chen24}). It thus seems likely that additional sources of dissipation are required to explain the high $\dot{P}$ of ASASSN-14ko, e.g., hydrodynamical drag as a star interacts with an accretion disc~\citep{linial24}.

\subsection{Comparison with other works}
Our analytical model (Section \ref{sec:lagrangian-model}) proposes that the expansion of the core consists of two parts, one of which is the ``background state,'' which consists of a sequence of quasi-steady and spherically symmetric configurations through which the star advances as the surface pressure declines (i.e., as the mass is removed). Because the pressure declines (the mass is removed) over a finite timescale, there is nonetheless residual kinetic energy contained in the star, the importance of which is then quantitatively recovered with linear perturbation theory applied to -- and with the eigenmodes of -- the asymptotic background state (achieved when the surface pressure goes to zero). This second part enables an understanding of the amplitude and timescale of the oscillations imparted to the star as a byproduct of the finite velocity of the core as it expands.

Our approach to determining the asymptotic background state is similar to the adiabatic mass loss model developed in~\cite{dai13}, who studied the response of stars undergoing dynamical mass transfer while in orbit around a compact object (see also~\citealt{hjellming87}, where the authors explored the adiabatic response and stability of various polytropic models undergoing mass loss). Specifically,~\cite{dai13} used conservation of entropy to compute the pressure in the interior of a star as a function of the core mass, $M_{\rm c}$. For a given mass loss, the central pressure for the corresponding $M_{\rm c}$ is then used as the boundary condition in the interior, in addition to setting $r(0)=0$ and $p(M_\mathrm{c})=0$, to solve the equation of hydrostatic balance and obtain the new radius as a function of enclosed mass. While we use a different boundary condition to emulate the decline in the surface pressure that would result from the stripping of mass from the surface layers of a star, our results for the asymptotic state -- and in particular the final average stellar density obtained as the pressure drops to zero -- agree both qualitatively and quantitatively with theirs, at least within the region of parameter space in which our stellar models overlap. The adiabatic mass-loss model has been used in other scenarios as well, e.g., binary star systems in which the donor undergoes rapid mass transfer, such that its response is always locally adiabatic \citep{ge10}, and for the early mass-transfer phases of a subgiant star orbiting an SMBH, before it undergoes a gravitational-wave driven inspiral~\citep{olejak25}.

\section{Summary and Conclusions}
\label{sec:summary}
We developed an analytical method for analyzing the response of stars to mass loss, ultimately to understand the survivability of stars in rpTDEs. The model is fundamentally simple, and describes the original star as a Bonnor-Ebert sphere that contains the inner $x\%$ of the original star by mass, with the remaining $\left(100-x\right)\%$ acting as the surrounding medium that -- for the background state -- retains pressure balance with the inner ``surviving core.'' The reduction in the pressure of the surrounding medium on $\sim$ the dynamical time of the star then mimicks the mass loss process, driving structural changes in the surviving core. 

Our main results are: 
\begin{enumerate}
    \item{The average density of high-mass stars, in particular those with masses $\gtrsim 0.7 M_{\odot}$, increases substantially for mass losses of $\lesssim 1-10\%$, and progressively with the mass lost, rendering them less susceptible to subsequent mass loss. The stars that experience the largest such increase in density have masses in the range $\sim 1.5-2 M_{\odot}$ (see Figure \ref{fig:mesa_stars_density_trend}). Contrarily, low mass stars progressively decrease in density with mass loss. An increase in the average density results in a decrease in the effective tidal radius of the star, making the star less susceptible to tidal stripping on subsequent encounters. This suggests that rpTDE candidates that exhibit successively dimmer outbursts with time (e.g., \citealt{wevers23, liu24}) are likely caused by high mass and evolved stars being repeatedly tidally stripped by an SMBH. Additionally, the star powering ASASSN-14ko -- if it is indeed fueled by accretion following the repeated tidal stripping of a star -- must have a mass $\gtrsim 1.5 M_{\odot}$ to have produced $\gtrsim 20$ flares.}
    \item{The change in the energy of a star in an rpTDE is predominantly determined by the difference in the binding energy of the core, prior to mass loss, and that of the original star. The work done by the expansion of the core is a small negative correction to the total energy. The tidal heating term is subdominant, and comparable in magnitude to the second order corrections to the gravitational and thermal energies. Tidal heating therefore does not excessively inflate the star or lead to runaway mass loss.}
    \item{The tidal interaction with an SMBH spins up the stellar core, and its rotational energy is second in the hierarchy of energy scales, the first being the difference in the binding energy of the core pre-mass-loss and the binding energy of the star as a whole. Hence, the different contributions to the total energy budget, in their relative order of importance are -- (1) the difference in the binding energy, $E_{\rm c,0}-E_\star$, (2) the rotational kinetic energy, (3) the $p-\mathrm{d}V$ work done by the expansion of the surface, and (4) the second order contributions from the kinetic, potential and internal energies (see Figure \ref{fig:ZAMS1-sph-energies} for an example).} 
\end{enumerate}

The model we developed to describe the mass loss process has broader applicability in the context of astrophysical phenomena involving mass ejection. As an example, it can be used to constrain the properties of outbursts of astrophysical systems that eject mass in response to the deposition of energy near the stellar surface, such as classical novae (\citealt{gallagher78}), Type IIn supernova progenitors~\citep{chevalier88, kochanek11,smith14}, and X-ray bursts~\citep{gottwald86}. Adopting a wind based mass loss prescription to model these scenarios, the perturbation analysis presented here can be generalized to model the nature of outbursts of such systems. Our model could also be used to study pulsating stellar systems such as $\delta-$Scuti stars, in which mass loss has been suggested to drive lithium depletion~\citep{russell95}, and luminous blue variables (LBVs), which are massive stars that exhibit mass loss driven outbursts on years-to-decades timescales~\citep{humphreys94,smith06}. 

\section*{Acknowledgements}
We thank Eliot Quataert for useful discussions, and the anonymous referee for useful comments and suggestions that improved the manuscript. A.B.~acknowledges support from NASA through the FINESST program, grant 80NSSC24K1548. E.R.C.~acknowledges support from NASA through the Astrophysics Theory Program, grant 80NSSC24K0897. Additional support for this work (A.B.) was provided by the National Aeronautics and Space Administration through Chandra Award Number 25700383 issued by the Chandra X-ray Observatory Center, which is operated by the Smithsonian Astrophysical Observatory for and on behalf of the National Aeronautics Space Administration under contract NAS8-03060. C.J.N.~acknowledges support from the Science and Technology Facilities Council (grant No. ST/Y000544/1) and from the Leverhulme Trust (grant No. RPG-2021-380). This research was supported in part by grant NSF PHY-2309135 to the Kavli Institute for Theoretical Physics (KITP).

\bibliographystyle{aasjournal}

\end{document}